\documentclass[twoside]{article}
\usepackage{iclr2026_conference,times}

\usepackage{natbib}
\usepackage{amsmath,amsfonts}
\usepackage{algorithmic}
\usepackage{algorithm}
\usepackage{array}
\usepackage[caption=false,font=normalsize,labelfont=sf,textfont=sf]{subfig}
\usepackage{textcomp}
\usepackage{stfloats}
\usepackage{url}
\usepackage{verbatim}
\usepackage{graphicx}
\usepackage{cite}
\usepackage{amssymb}
\usepackage{enumitem}
\usepackage{booktabs}
\usepackage{bm}
\usepackage{tabularx}
\usepackage{subcaption}
\usepackage{xcolor}
\usepackage{siunitx}
\usepackage{float}
\usepackage{listings}
\usepackage{setspace}
\usepackage{xspace}

\definecolor{mutedblue}{RGB}{70,100,200}
\usepackage[colorlinks=true,
            citecolor=mutedblue,
            linkcolor=black,
            urlcolor=mutedblue]{hyperref}

\usepackage[group-separator={,}]{siunitx}

\iclrfinalcopy
\begin{document}

\title{\vxp: A Vision Agent for End-to-End Medical Image Analysis}

\author{Andrew Hoopes~\textsuperscript{1,2},~
Neel Dey~\textsuperscript{1,2,3},~
Victor Butoi~\textsuperscript{1},~
John ~Guttag~\textsuperscript{1},~
Adrian V.~Dalca~\textsuperscript{1,2,3} \\
\vspace{0.5em}
{\small
\textbf{1.}~Massachusetts Institute of Technology,
\textbf{2.}~Massachusetts General Hospital,
\textbf{3.}~Harvard Medical School
}}

\newcommand{\subpara}[1]{\vspace{0.1em}\noindent\textbf{#1.}\hspace{0mm}}
\newcommand{\vxp}{VoxelPrompt\xspace}

\newcommand*{\real}{\mathbb{R}}
\newcommand*{\task}{\tau}
\newcommand*{\taskset}{\mathcal{T}}
\newcommand*{\prompt}{p}
\newcommand*{\invols}{\mathcal{V}}
\newcommand*{\environ}{\Omega}
\newcommand*{\library}{\mathcal{N}}
\newcommand*{\agent}{\alpha}
\newcommand*{\enc}{m_\text{enc}}
\newcommand*{\gen}{m_\text{gen}}
\newcommand*{\instr}{\varphi}
\newcommand*{\code}{c}
\newcommand*{\state}{\mu}
\newcommand*{\encvols}{\mathcal{E}}
\newcommand*{\outvols}{\mathcal{W}}
\newcommand*{\cat}{\shortparallel}

\maketitle
\thispagestyle{firstpage}

\begin{abstract}
We present \vxp, an end-to-end image analysis agent that tackles free-form radiological tasks. Given any number of volumetric medical images and a natural language prompt, 
\vxp integrates a language model that generates executable code to invoke a jointly-trained, adaptable vision network. This code further carries out analytical steps to address practical quantitative aims, such as measuring the growth of a tumor across visits. The pipelines generated by \vxp automate analyses that currently require practitioners to painstakingly combine multiple specialized vision and statistical tools. 
We evaluate \vxp using diverse neuroimaging tasks and show that it can delineate hundreds of anatomical and pathological features, measure complex morphological properties, and perform open-language analysis of lesion characteristics. \vxp performs these objectives with an accuracy similar to that of specialist single-task models for image analysis, while facilitating a broad range of compositional biomedical workflows. 
\end{abstract}

\begin{figure*}[!ht]
    \centering
    \includegraphics[width=\textwidth]{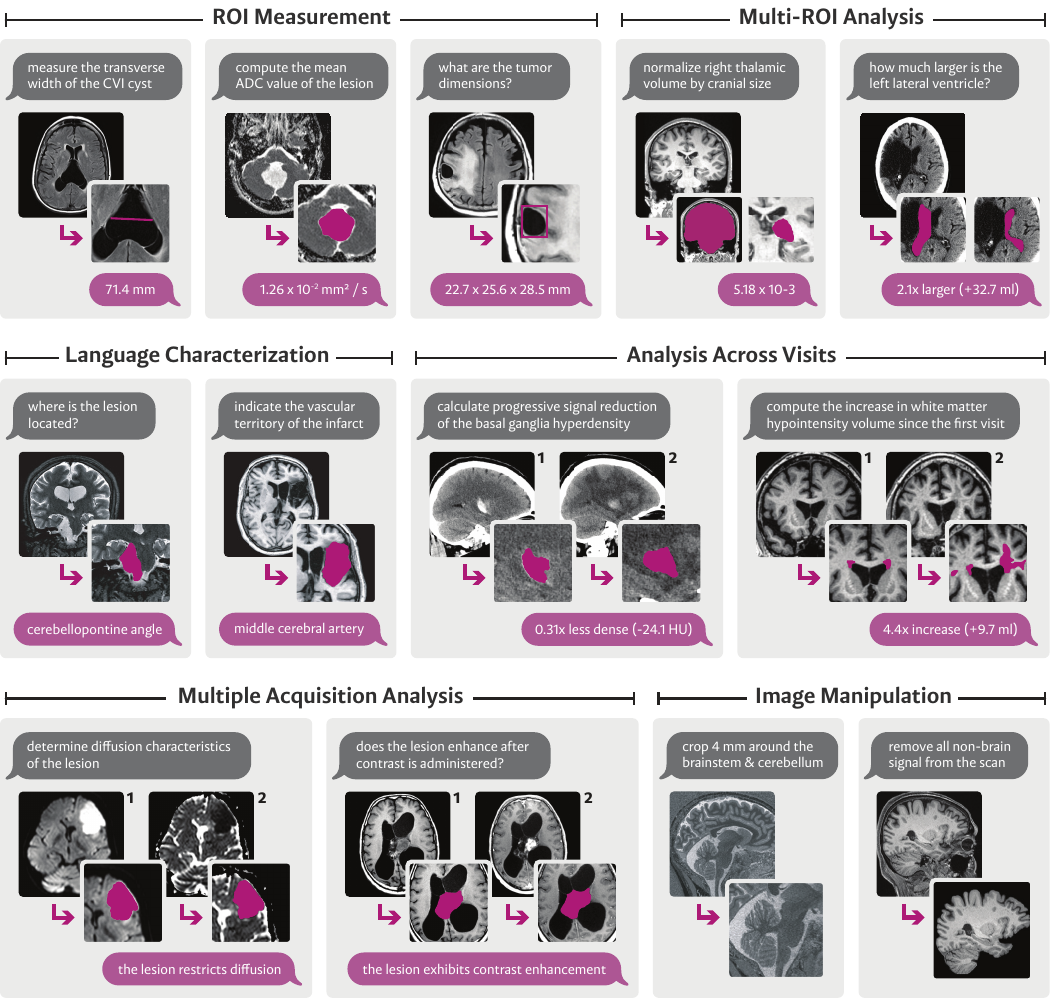}
    \caption{Illustrative examples of \vxp capabilities, each showing the input prompt (gray) and volumes with \vxp's predicted annotations and language responses (purple).
    }
    \label{fig:showcase}
\end{figure*}

%
%

\section{Introduction}

Clinicians and scientists routinely pose complex questions involving specific targets in medical imaging that extend well beyond simple segmentation or classification tasks. These questions involve multi-step efforts to track the evolution of a particular pathology over many scans, quantify subtle asymmetries of a specific anatomy, or integrate information from multiple acquisitions. 

As a detailed example, consider tracking the growth of a specific lesion over time in a patient with multiple abnormalities. After image pre-processing, the first challenge is segmenting \textit{only} the specific lesion of interest. Available tools rarely generalize to diverse, real-world lesion types, and even those that do offer no way to identify a specific lesion using natural language descriptors (e.g., by anatomical location, size, or intensity). Additionally, current tools do not typically accommodate a flexible number of acquisitions from a scan session. As a result, the user must choose a single suitable scan, develop a custom pipeline to programmatically select the target lesion, repeat the process for later scans, and then compute the desired downstream metrics to track changes.

The example above illustrates a fundamental barrier in integrating AI in real imaging workflows. 
While existing tools perform well on specific segmentation or classification targets~\citep{billot2023synthseg,isensee2025nninteractive}, they are specialized to their intended use cases and cannot be used directly to perform broader, integrated analyses that require executing multiple steps.
This task-level specialization limits the adoption of AI in radiology, leading to practitioners with complex radiological questions needing to manually chain together multiple fragile, task-specific models and develop extensive post-processing and metric-extraction workflows for each new study.

\vxp is fundamentally different in functionality and design from existing medical image analysis systems. In \vxp, we jointly train a language model agent and vision network from scratch to generate and execute \textit{end-to-end} image analysis workflows. Given a task described in natural language, the agent iteratively orchestrates a sequence of instructions as executable code. The dynamically evaluated instructions generate spatial features (e.g., segmentations) using the vision network, incorporate natural language responses, and access a library of functions to compute and provide quantitative outputs. Through diverse output modalities, \vxp can segment and localize user-specified anatomical and pathological regions of interest, calculate measurements that relate multiple scans to one another, and perform biomedical characterization~(Figure~\ref{fig:showcase}).

\looseness=-1
We make several technical contributions to realize \vxp's capabilities for real-world medical imaging aims. 
Our convolutional vision network enables fine-grained, language-controlled visual analysis by integrating jointly-trained language model embeddings as conditioning. To also support multi-acquisition and longitudinal studies~\citep{reuter2012within}, the vision network uses attention to process volumetric features across sequences of any length. Further, unlike typical models tied to fixed channels and voxel spacing~\citep{zhang2024foundation}, \vxp operates on variable-sized inputs at native resolution. This native processing yields substantial memory and runtime efficiency, enabling the joint training of vision and language components on large neuroimaging volumes on standard GPUs without prior resampling. Lastly, we facilitate robustness to acquisition type as well as anatomical and pathological variation by creating and training on a large neurological dataset combining public cohorts, new annotations of unlabeled pathological volumes, and simulated lesions.

\looseness=-1
We focus on brain imaging and show that \vxp enables end-to-end analysis on nuanced and diverse tasks covering a wide range of MRI and CT acquisitions, anatomies, and diseases. Quantitatively, we show that a single \vxp model captures, and often exceeds, the individual accuracy and capabilities of many single-task specialist neuroimaging baselines, while retaining unique language prompted flexibility. These results highlight \vxp's promise as a foundation for tackling diverse and complex radiology workflows.

%
%

\section{Related Work}

\subpara{Brain Region Analysis} Widely-used neuroimage analysis pipelines typically delineate regions and quantify their size, shape, composition, and change over time~\citep{fischl2012freesurfer,jenkinson2012fsl}. Modern approaches train networks to segment anatomical and pathological structures, including cerebral subregions~\citep{billot2023synthseg,henschel2020fastsurfer}, vessels~\citep{hilbert2020brave,livne2019u}, and lesions~\citep{hssayeni2020intracranial,liu2021deep}. While performant, these networks generally work for fixed segmentation targets and require significant human involvement for analyzing data and deriving downstream ROI measures. \vxp aims to match or outperform these methods in segmentation accuracy, while tackling a wider range of targets, enabling flexible specification of target regions, and facilitating end-to-end workflows.

\subpara{Learning Across Medical Imaging Tasks}
Recent medical imaging methods aim to improve performance by exploiting shared representations across diverse segmentation, classification, registration, and statistical modeling objectives in a single framework~\citep{elmahdy2021joint,graham2023one,tellez2020extending,liu2025modality, czolbe2023neuralizer}. 
Broad, segmentation-focused tools, like interactive or in-context segmentation models, can adapt to specific biomedical targets, prompted by partial image annotations~\citep{cheng2023sam,luo2021mideepseg,ma2024segment,wong2023scribbleprompt} or example image-segmentation pairs~\citep{min2021metaicl,xie2021explanation,butoi2023universeg,ouyang2022self,rakic2024tyche,roy2020squeeze}. However, these multi-task models do not aim to address a complete analytical pipeline and can require finetuning in real scenarios. In contrast, \vxp integrates supervision from many tasks to create computational workflows, where multiple components interact to carry out requested analyses.

\subpara{Medical Vision-Language Models}
Vision-language models (VLMs) trained on large-scale biomedical image-caption datasets~\citep{johnson2019mimic,lin2023pmc,zhang2023biomedclip} can facilitate biomedical visual question-answering~\citep{chen2023medblip,chen2023generative,zhang2023biomedclip,zhang2023pmc} and clinical report generation~\citep{bannur2024maira,wang2023r2gengpt,wang2023metransformer}.
However, current biomedical VLMs remain largely limited to narrow-domain, text generation tasks, and do not capture the quantitative metrics required in real-world clinical imaging workflows. In contrast to current vision-language models that produce text outputs in a black-box manner, \vxp explicitly produces code for all relevant intermediate outputs and a traceable sequence of operations. This provides analytical transparency for high-stakes applications. Also, unlike existing models, the \vxp operations involve explicit vision operations to compute and present images depicting the essential intermediate features. Finally, aside from few recent works~\citep{chen2023medblip,chen2023generative,liu2023t3d,wu2025towards,zhou2024generalist}, most models are trained exclusively on two-dimensional image slices, often X-rays, making them inappropriate for MR and CT imaging. \vxp is instead trained directly at native acquisition resolution, enabling it to process 3D volumes. 

\subpara{Language Models as Agents} Recent efforts extend large language models beyond plain text prediction into agents capable of planning and executing actions for computational tasks. Often, these generate code~\citep{gupta2023visual,ke2025explain} that call external APIs for mathematical computation~\citep{ruan2023tptu,gou2023tora}, image analysis~\citep{subramanian2023modular,suris2023vipergpt,yang2023mm}, scientific discovery~\citep{bran2023chemcrow,boiko2023emergent}, and more. Adaptive, feedback-driven agents address complex and dynamic problems by iteratively planning, executing, and interpreting intermediate outcomes rather than predicting entire action sequences at once~\citep{huang2022inner,rana2023sayplan,wang2023describe,wang2023voyager,yao2022react,zhu2023ghost}. Building on this idea, \vxp trains an adaptive agent that interacts with a library of processing functions. Unlike other methods, \vxp jointly trains an adaptable vision network to guide image processing. Recent work in medical imaging~\citep{li2024mmedagent} trains an agent to select from a set of pretrained, task-specific tools, but it does not execute downstream operations or leverage flexible language prompting to distinguish ROIs with specific characteristics as in \vxp.

%
%

\section{Methods}
\label{sec:methods}

\subsection{Modeling Details}
\label{subsec:modeling_details}

\begin{figure*}[t]
    \centering
    \includegraphics[width=0.95\textwidth]{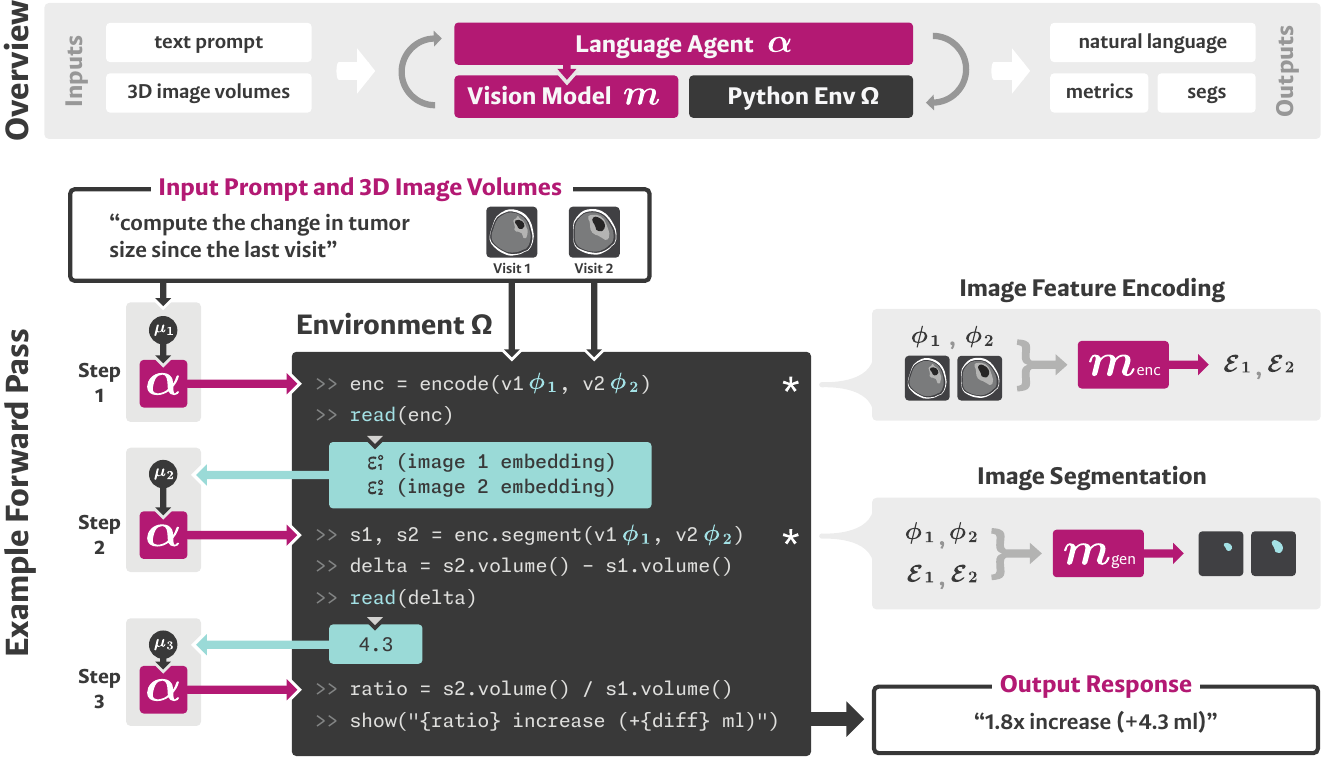}
    \caption{\textbf{Top:} \vxp takes a text prompt and volumes as input to a trainable agent~$\agent$. The agent iteratively produces executable code in a Python environment~$\Omega$, which controls a jointly-trained vision model $m$.
    \textbf{Bottom:} To solve an example language-prompted task,~the agent $\agent$ interprets execution outcomes~$z$ (blue) to guide subsequent instruction prediction across multiple steps. To perform vision operations, such as volume encoding or generation,~$\alpha$ employs vision networks~$\enc$ and~$\gen$, which are manipulated by image-specific latent instruction embeddings~$\phi$.}
    \label{fig:method-example}
\end{figure*}

\vxp processes volumes~$\invols$ in response to a text prompt~$\prompt$. A language model agent $\agent$ translates the prompt into Python code executed in a persistent environment $\Omega$, invoking actions involving mathematical computation, morphological operations, and interface interaction. A core function set runs a jointly trained vision network directed by the agent to perform vision operations. Figure~\ref{fig:method-example} summarizes this framework and outlines an example use-case. We provide low-level implementation details in Appendix~\ref{appendix-implementation}.

\subpara{Agent} The agent~$\agent$ produces code iteratively, with each step building on outcomes of prior actions. At step~$i$, it generates executable code~$\code_i = \agent(\state_i)$ based on a state representation~$\state_i \in \real^{\ell,d}$ with sequence length~$\ell$ and embedding dimension~$d$. The code runs in environment~$\Omega$, which preserves variables across steps. Intermediate results from~$\Omega$ can be read and embedded into a representation~$z_i$ and incorporated into the next state $\mu_{i+1} = \mu_i \cat z_i$, where~$\cat$ denotes sequence concatenation. The initial state~$\state_1$ encodes the prompt~$\prompt$ and acquisition date metadata for each volume~$v \in \invols$. This loop of code generation, execution, and feedback continues until a stopping code signals completion.

\subpara{Vision Network} Several functions in environment~$\Omega$ invoke a shared convolutional vision network, consisting of an input feature encoder $\encvols = \enc(\invols, \phi)$ and a volume generator~$\outvols = \gen(\encvols, \phi)$. The latent embeddings~$\phi$ are produced by the agent, passed as arguments to the functions, and condition the vision network for a specific goal. For example, they might direct the vision network to segment the edema around the lesion in the frontal lobe.  Vision network outputs $\encvols$,~$\outvols$ can then be further processed by downstream actions to execute the user prompt.  

\looseness=-1
\subpara{Volume Interaction}
The vision subnetworks share information across an arbitrary number of input volumes using an attention mechanism.
Each input volume~$v$ is processed by~$\enc$ (or its encoding~$\encvols_v$ processed by~$\gen$), producing individual \textit{streams} of intermediate activations from each volume that interact with each other at each layer.
Specifically, for voxel features~$a_s \in \real^c$ in volume stream~$s$, we concatenate~$a_s$ with stream-specific~$\phi_s$, and use a fully-connected layer to yield~$a'_s \in \real^c$. We then stack corresponding voxel representations~$a'_s$ across~$S$ streams to construct~$A' \in \real^{S,c}$. We interact streams in~$A'$ using attention with dimension~$b$: $B = f (\text{softmax} (Q K^T b^{-1/2})V) + A,$ 
where~$Q, K, V \in \real^{S,b}$ are learnable linear transformations of~$A$, and fully-connected layer~$f$ projects the output to~$\real^{S,c}$. We then separate $B$ into stream-specific voxel features for each volume.

\subpara{Native Space Processing}
Volumetric image formats define a world-coordinate transform specifying in-plane voxel spacing~$x$ and inter-slice spacing~$y$, with anisotropy ratio $r = y /x$. Standard tools often resample images to isotropic resolution, which greatly inflates data size for thick-slice acquisitions. Instead, we implement a vision network that operates in native voxel resolutions by tracking and updating spacings throughout the multi-scale hierarchy. In the downsampling arm, following resolution level $n$, the target in-plane spacing is set to $x_{n+1} = 2 x_n$, while the target slice spacing is updated to $y_{n+1} = x_{n+1}$ only when $r_n \leq 2$. In the upsampling arm, voxel spacings are inferred from the skip connections. During stream interaction, volume features are resampled to a common geometry, then returned to their previous space.

\subpara{Supervised Training}
We jointly train the language model~$\agent$ and vision network from scratch on a curated, diverse task set ~$\taskset$ (Section~\ref{tasks}). Each task~$\task \in \taskset$ is paired with target (ground-truth) code~$\code^*$ that carries out the task objective, as illustrated in Figure~\ref{fig:method-example}. At each training step, we sample~$\task \sim \taskset$, generate a prompt~$\prompt$, and sample input volumes~$\invols$ with ground-truth outputs~$\outvols^*$. The training loss is $\mathcal{L}_{ce}\big(P(\code), \code^*\big)
+ \lambda ~
\sum_{j=1}^{|\outvols|} \mathcal{L}_{img}\big(\outvols_j, \outvols_j^*\big)
,
$
where~$P(\code)$ is the language model output, volumes~$\outvols$ are generated by the vision networks while executing~$\code^*$, $\mathcal{L}_{ce}$ is cross-entropy, and $\mathcal{L}_{img}$ compares predicted and target volumes (using soft Dice loss for segmentation).

%
%

\subsection{Training Tasks and Data Design}
\label{tasks}

We construct and curate a dataset~$\taskset$ of brain imaging tasks, designing new task formulations and labels across a wide range of image acquisitions, segmentation protocols, and annotation types. We use this dataset to both train and evaluate \vxp in the joint prediction of analytical instructions, spatial delineations, and natural language descriptions. We include a set of clinically-oriented objectives, which are broadly categorized as either ROI processing or pathology description tasks. For each task, we create ground-truth code~$\code^*$, used in training and evaluation. Additional training data details are described in Appendix~\ref{appendix-training-data}.

\subpara{Training Code for Quantitative ROI Processing}
Quantitative processing tasks involve image feature segmentation, optionally followed by downstream steps to compute ROI measures. We include a core segmentation task for all structures and pathology classes in our dataset. Downstream processing tasks use predicted segmentations, sometimes in conjunction with the input volumes. For example, some tasks involve removing, extracting, or cropping the field of view~(FOV) around a segmented region. Others use segmentations to compute ROI-specific statistics of image signal intensities (e.g., mean intensity). Morphological tasks analyze ROI shape and compute total volume, bounding box dimensions, or the maximum height, width, and depth of a segmented structure. We also include tasks that compute and compare such metrics across multiple segmentations. For example, longitudinal tasks measure change in ROI properties across a series of scan sessions, and multi-region tasks compare metrics from different ROIs in a single scan session. To support these applications, ground-truth code~$\code^*$ specifies a sequence of functions that predict the required segmentations, compute relevant metrics, and format the results into an output message (Figure~\ref{fig:method-example}).

\subpara{Training Code for Question Answering}
We also train on question-answering tasks, where \vxp produces natural language responses from a combinatorially large set of possible answers. For example, some tasks involve classifying lesion signal intensity as hyperintense, hypointense, or isointense relative to surrounding tissue, while others require identifying anatomical location. 
Certain tasks integrate information across multiple images, for example, detecting restricted diffusion from paired DWI and ADC maps, or assessing post-contrast lesion enhancement. We handcraft a target language response for all possible answers. For each task, we construct the ground-truth instruction code~$\code^*$ with functions to (1) encode the input volumes, (2) read the encoded volume features, and (3) output a text response with the correct natural language answer.

\subpara{Training Prompt Synthesis}
\label{prompt-synth}
We synthesize a combinatorially diverse set of prompts for training. 
For each task~$\task$, we define a set of prompt templates~$\mathcal{P}_\task$ containing placeholders to accommodate multiple words, terminologies, and phrases with similar meanings. We compile a list of interchangeable text~$\mathcal{C}_k$ for each placeholder~$k$. To generate a prompt~$\prompt$ for task~$\task$, we sample a template from $\mathcal{P}_\task$, then fill each placeholder~$k$ with text sampled from~$\mathcal{C}_k$. Placeholders may themselves contain other placeholders, making the process recursive. This yields a diverse distribution of prompts spanning clinical and imaging terminology, as well as variations in tense, syntax, and word choice.

\subpara{Training Images and Segmentations}
\label{images}
We assemble and annotate a collection of  6,925 3D brain MRI and CT scans from 15 public datasets, comprising 185 bilateral anatomical structures and 14 pathology classes, focusing on a breadth of imaging types, regions of interest, and tasks. The MRI sequences span T1w, T2w, FLAIR, PD, GRE, and DWI with various scan resolutions. The subjects are split into 4,852 training, 213 validation, and 1,860 test volumes. Anatomical segmentations are derived from established pipelines~\citep{fischl2012freesurfer,greve2021deep,hoopes2022synthstrip}, atlas annotations~\citep{adil2021high,pauli2018high}, manual corrections, and manual labeling of additional structures in a small set of images, yielding high-quality whole-brain labels across multiple cohorts.

To capture diverse pathologies, we integrate expert-annotated lesions from BraTS, ISLES, ATLAS, and WMH~\citep{baid2021rsna,hernandez2022isles,liew2022large,kuijf2019standardized}, covering gliomas, edema, infarcts, and white matter hyperintensities. We further compile rare cases from \textit{Radiopaedia} and manually delineate infarcts, arachnoid and epidermoid cysts, papillomas, and many others. These new annotations also include sub-components like edema, enhancing tissue, and heterogeneous intra-lesion features. Finally, we augment the dataset with a conditional synthesis procedure that generates diverse lesions in healthy brains, broadening the distribution of pathological presentations (Appendix~\ref{appendix-lesion-synth}). To support analysis of lesion characteristics, we annotate each lesion with its anatomical location, intensity profile, size, and position relative to surrounding structures, and, when applicable, indicators of diffusion restriction or post-contrast enhancement.

%
%

\section{Experiments}
\label{experiments}

\vxp addresses non-standard, open-ended workflows rather than a single fixed task. Its evaluation, therefore, requires a diverse set of complementary experiments. We first present experiments that evaluate \vxp's ability to generate and execute accurate end-to-end brain analyses across several representative practitioner use-cases. We then present analyses and ablations of modeling decisions. We provide further experimental and test data details in Appendix~\ref{appendix-experiments} and include additional results demonstrating disease characterization performance in Appendix~\ref{appendix-classification-results}.

\begin{figure}[!t]
    \centering
    \includegraphics[width=\linewidth]{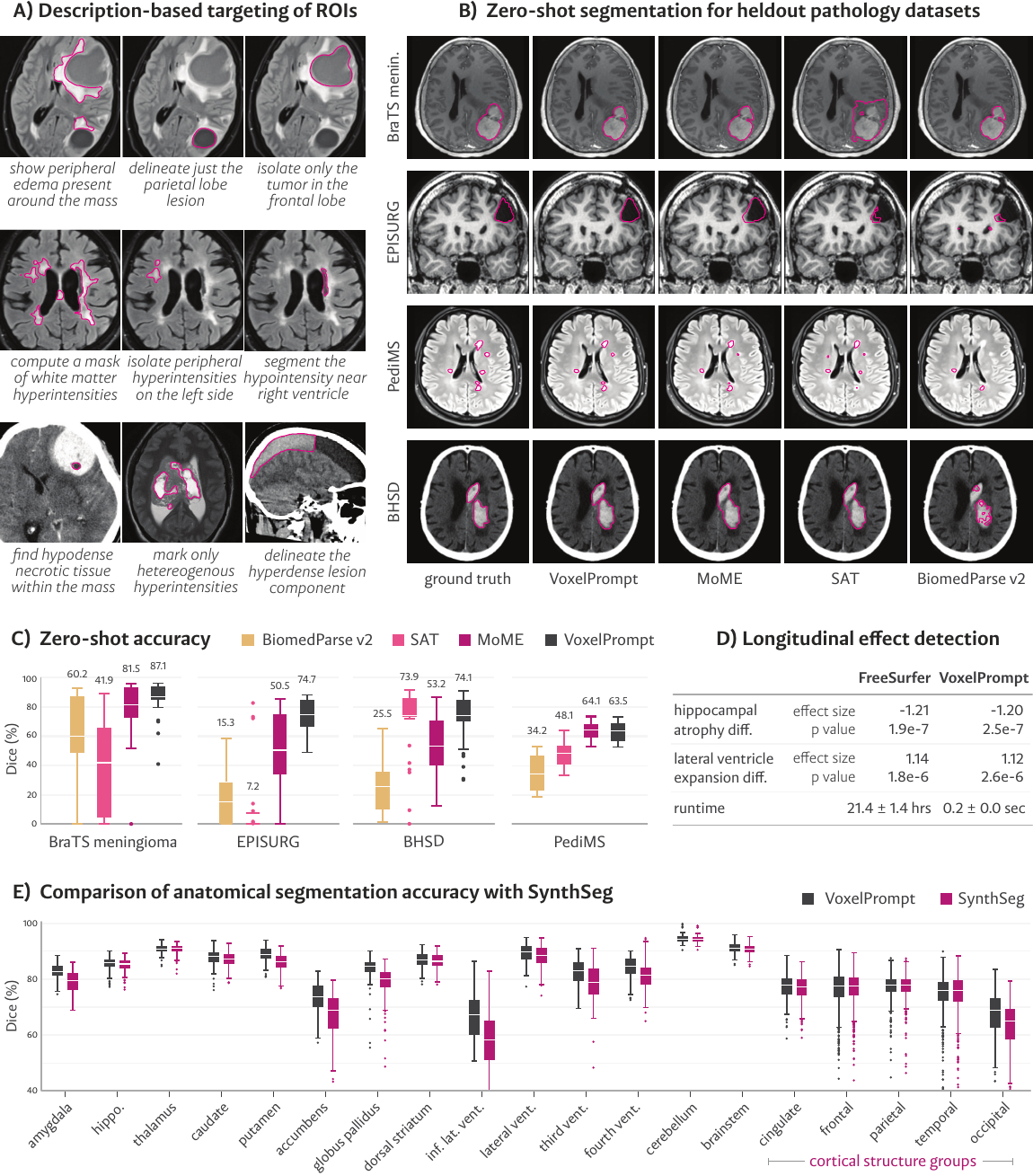}
    \caption{\textbf{\vxp performance.} \textbf{(A)} Free-form text prompts, shown below each image, guide \vxp to perform targeted analysis and delineation of nuanced, context-specific image regions, even in scans with multiple lesions. \textbf{(B, C)} On unseen datasets with diverse brain abnormalities, \vxp is the only method achieving consistently high-quality results both qualitatively and quantitatively. \textbf{(D)} Compared to longitudinal FreeSurfer, \vxp achieves the same effect size in distinguishing Alzheimer’s disease from controls with a $10^{5}\times$ faster runtime. \textbf{(E)} \vxp outperforms the state-of-the-art specialist model (SynthSeg) on whole brain segmentation.
    }
    \label{fig:main_results}
\end{figure}

\subsection{Brain Image Analysis} \label{subsec:brain-image-analysis}

\subpara{\textit{Ad hoc} Neuroimaging Workflow Generation} 
Figure \ref{fig:showcase} shows that a single \vxp model can execute a wide range of workflows on held out test data, including localizing brain anatomy and pathology regions, extracting intensity metrics and morphology measures within user-specified ROIs, and masking or cropping tissues for focused visualization. The model can compute and compare metrics across ROIs, such as hippocampal asymmetry, normalized subcortical volumes, and acute versus chronic hemorrhage components, as well as track longitudinal changes such as tumor size across scans. By integrating multiple acquisitions, \vxp can further characterize lesion locations and tissue properties, such as diffusion restriction or post-contrast enhancement. Figure \ref{fig:main_results}A shows that \vxp facilitates fine-grained specificity by supporting flexible, language-guided analysis, such as isolation or differentiation of lesions in multifocal disease based on signal intensity, size, relative position, or anatomical context (e.g., hemisphere, lobe, etc.). Examples in Figure~\ref{fig:showcase} reflect common practitioner use-cases, but many are qualitative as no benchmark dataset currently exists to evaluate free-form workflow generation outcomes for complex pathology analyses.

\subpara{Text-prompted Zero-shot Lesion Segmentation}
\label{results:lesion_seg}
Zero-shot brain lesion segmentation enables medical practitioners and researchers to rapidly localize and quantify pathologies without requiring a disease-specific model.
We evaluate \vxp's zero-shot segmentation capabilities on entirely unseen abnormality datasets using a dataset-specific prompt: ``segment the $\langle ROI \rangle$," where $\langle ROI \rangle$ is the target lesion type or informative description.
We benchmark our approach against multi-dataset foundation models that include brain pathology segmentation as training tasks. 
These include volumetric BiomedParse v2~\citep{zhao2024biomedparse,zhao2025biomedparsev} and SAT~\citep{zhao2025large}, both of which use text prompts to target abnormalities. Most other existing text-prompted vision models, or vision–language models (VLMs), are limited to question-answering tasks, and do not perform quantitative analyses such as segmentation, thereby precluding them as baselines. We also use MoME~\citep{zhang2024foundation}, a recent generalizable brain abnormality segmentation model. Our evaluation suite spans diverse targets: 30 meningiomas from BraTS-MEN~\citep{labella2024multi}, 9~pediatric MRIs of multiple sclerosis lesions from PediMS~\citep{popa2025pedims}, 35 resection cavities from EPISURG~\citep{perez2020simulation}, and 36 hemorrhages from BHSD~\citep{wu2023bhsd}. 
\vxp has not been trained on these datasets, allowing for zero-shot performance assessment.

Figures~\ref{fig:main_results}B and C demonstrate that \vxp is the only method that achieves consistently high performance across all lesion target types, and on average achieves the highest Dice score. We find that no baseline achieves generalization across abnormalities, and the second-best method varies from dataset to dataset. Quantitatively, \vxp achieves a mean 12.53 Dice points higher than MoME, the overall second-best method. Appendix Figure~\ref{fig:appendix-ranked} shows per-subject performances.

\subpara{Whole Brain Anatomical Segmentation}
\label{results:whole_brain_results}
We evaluate \vxp's ability to segment diverse neuroanatomical targets. Since many brain structures are bilateral, we prompt \vxp to ``segment the left and right  $\langle ROI \rangle$" to generate joint segmentations, where applicable. We compare against the widely used state-of-the-art SynthSeg v2~\citep{billot2023synthseg} method for multi-class anatomical segmentation, which generalizes across the diverse acquisition contrasts exhibited in our image dataset. We use a structural MRI test set of 108 unseen volumes, which span various tissue contrasts and contain ground-truth segmentations for the 45 structures predicted by SynthSeg.

Figure~\ref{fig:main_results}E shows that \vxp significantly outperforms SynthSeg ($p < 0.05$) on 23/45 ROIs, with a mean Dice improvement of~$+1.1 \pm 2.3$\% over all structures. 
While \vxp achieves a modest improvement, we emphasize that our main goal is not to outperform established tools for segmentation sub-tasks, but rather to provide reliable anatomical segmentations while retaining \vxp's unique flexibility of natural language prompting for workflows.

\begin{figure}[!t]
    \centering
    \includegraphics[width=0.98\linewidth]{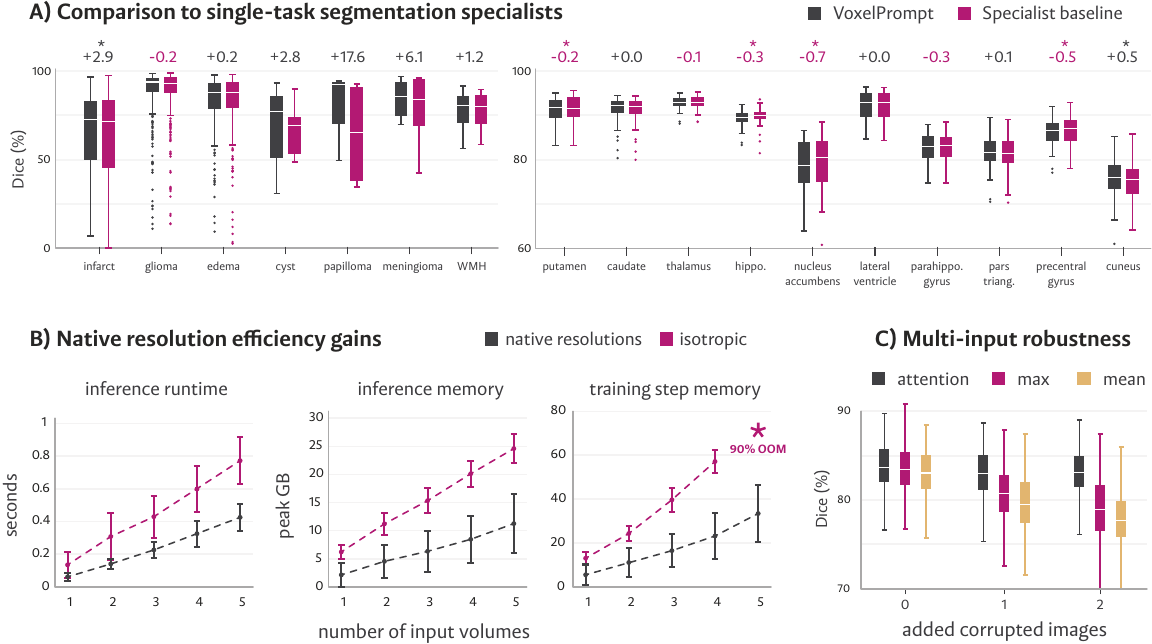}
    \caption{\textbf{Ablations and analyses}. \textbf{(A)} A single \vxp model trained jointly on all tasks matches or exceeds task-specific models for both lesions (left) and anatomy (right). Asterisks indicate statistically significant differences. \textbf{(B)} Our proposed native-resolution convolutions are more efficient in runtime and memory than isotropic resampling. \textbf{(C)} Our attention mechanism for multi-input volume interaction is more robust to image corruptions compared to max and mean reductions.}
    \label{fig:ablation}
\end{figure}

\looseness=-1
\subpara{Longitudinal Analyses}
\label{longitudinal_results}
 Figure~\ref{fig:showcase} shows qualitatively that \vxp performs volumetric analyses across time for various pathologies. Here, we quantify performance for longitudinal analyses of \textit{anatomical} structures, a core component of large neuroimaging studies. We aggregate 100 subjects with two MRI sessions separated by two years from the ADNI dataset~\citep{weber2021worldwide}, equally split between controls and Alzheimer's disease (AD) subjects. We assess \vxp's off-the-shelf ability to measure the change in AD-affected structures over time, and use that to distinguish controls from AD subjects. Specifically, we compare effect sizes and runtime against longitudinal FreeSurfer~\citep{reuter2012within}, the widely used standard for multi-session analysis. As demonstrated in Figure~\ref{fig:main_results}D, \vxp can detect well-established AD-related effects, such as increased hippocampal atrophy and increased ventricle volume expansion over time, with a similar effect size to FreeSurfer, while offering a dramatic~$3.8 \times 10^5$ speedup in runtime.

\subsection{Ablations and Analyses} \label{subsec:ablations}

\subpara{Multi-Task Training}
We test whether the proposed \textit{single} \vxp model trained jointly on multiple tasks can match the performance of single-task specialist networks. We optimize individual, label-specific segmentation networks with the same architecture as the \vxp vision network, for a subset of distinct segmentation tasks, using a soft Dice loss. 
Since optimizing a specialized baseline for each ROI in our training dataset is computationally prohibitive, we select a subset of 10 anatomical and 7 pathology targets spanning diverse shapes and locations. In total, the resulting evaluation subset encompasses 638 held-out subjects.

Figure~\ref{fig:ablation}A shows that \vxp performance is on par with~($p > 0.05$) or exceeds~($p < 0.05$) the performance of~$13/17$ single-task specialists. 
The mean Dice difference relative to the specialists is $+4.3 \pm 5.7 \%$ for pathology targets and $-0.1 \pm 0.3 \%$ for anatomical structures.
This shows that multi-task training in \vxp rivals specialist models, while offering substantial improvement for brain abnormality segmentation, especially for variable lesions and limited data.

\subpara{Native Resolution Efficiency}
We evaluate the efficiency gains of the proposed native-resolution vision network by comparing the processing of images at their native resolution to the standard approach of conforming all inputs to a 1 mm$^3$ isotropic geometry~\citep{billot2023synthseg}. For both approaches, we measure inference runtime, peak GPU memory during inference, and training memory incurred during the backward pass of a single optimization step using the \vxp model. Image geometries are drawn from a distribution reflecting those encountered in a single clinical scan session (Appendix~\ref{appendix-scan-resolutions}), and results are averaged over 500 samples. To assess scalability, we repeat the experiment with increasing numbers of input images per sample.

Figure~\ref{fig:ablation}B shows that, averaged across all numbers of input volumes, native-resolution processing achieves a~$2\times$ reduction in inference runtime and a~$2.4\times$ memory reduction compared to isotropic resolution conformation. Isotropic resampling in training incurs~$2.2\times$ higher memory costs, rendering it challenging to train a multi-modal model that accepts only isotropic inputs on standard hardware. For example, with 5 input volumes, frequently present in a longitudinal MRI series, isotropic inputs cause out-of-memory errors on $90\%$ of sampled batches on an 80 GB GPU.

\subpara{Mechanisms of Stream Interaction}
We compare our attention-based interaction module with existing non-parametric alternatives~\citep{butoi2023universeg} that interact features across image inputs using mean or max reductions. To isolate the effect of interaction, we train vision-only models using three mechanisms: attention, mean, and max reduction (Appendix~\ref{appendix-interaction-results}). 
We train models using synthetic multi-contrast brain images generated from 360 MRIs using a domain randomization pipeline~\citep{gopinath2024synthetic}. Synthetic inputs enable us to explicitly test accuracy gains from multi-contrast integration. We also measure robustness to groups of images of variable quality by synthetically corrupting a subset of images in a group. 
We evaluate using 500 synthetic brains with arbitrary contrasts, measuring average Dice over 35 anatomical labels. We find that all three mechanisms achieve similar overall accuracy improvements as the number of uncorrupted inputs is increased (see Appendix Table~\ref{tab:interaction-variants}). 
However, as shown in Figure~\ref{fig:ablation}C, attention interaction is markedly more robust to image quality: when including corrupted inputs, Dice degrades by only~$0.6 \pm 3.4\%$, compared to~$4.6 \pm 4.5\%$ and~$5.4 \pm 4.0\%$ for max- and mean-reduction, respectively.

%
%

\section{Discussion}

\looseness=-1
\subpara{Limitations and Future Work} The \vxp vision and language networks are trained from scratch on simulated user prompts generated from templates, which limits their utility when given entirely unseen prompts. This limitation can be addressed by constructing more diverse datasets containing a broader range of tasks, either through simulations or language instructions from real users. While \vxp was trained on a combination of public brain imaging datasets and images we aggregated and annotated from the open Internet, its training set can be extended further by inferring tasks from clinical images and their associated reports, which might better cover the complex edge-case pathologies seen in practice. Additionally, instead of training the language model from scratch, a lightweight pre-trained language model with broad programming and natural language knowledge could be finetuned to support generalization to new tasks. We believe that our training strategy is generic and could be productively applied to radiology beyond neuroimaging.

\subpara{Conclusions} We introduced \vxp, a vision-language system that can address radiological aims not possible with existing methods as well as tasks that today require a multitude of specialized models and extensive manual user work. We demonstrated that agent-based \vxp accurately solves a broad spectrum of neuroimaging tasks involving end-to-end image analysis. Moreover, it provides transparent execution steps that can provide users with confidence in its results. We anticipate \vxp's use in projects, \textit{ad hoc} studies, and clinical pipelines, empowering biomedical users to adopt AI into their medical imaging workflows.

\clearpage
\newpage

\section*{Acknowledgements}

This work is supported in part by the National Institute of Biomedical Imaging and Bioengineering~(R01~EB033773, T32~EB001680), the Harvard MIT Neuroimaging Training Program, the National Science Foundation Graduate Research Fellowships Program, Quanta Computer Incorporated, and Felicis Ventures.

\bibliographystyle{iclr2026_conference}
\bibliography{main}

\begin{thebibliography}{85}
\providecommand{\natexlab}[1]{#1}
\providecommand{\url}[1]{\texttt{#1}}
\expandafter\ifx\csname urlstyle\endcsname\relax
  \providecommand{\doi}[1]{doi: #1}\else
  \providecommand{\doi}{doi: \begingroup \urlstyle{rm}\Url}\fi

\bibitem[Adil et~al.(2021)Adil, Calabrese, Charalambous, Cook, Rahimpour, Atik, Cofer, Parente, Johnson, Lad, et~al.]{adil2021high}
Syed~M Adil, Evan Calabrese, Lefko~T Charalambous, James~J Cook, Shervin Rahimpour, Ahmet~F Atik, Gary~P Cofer, Beth~A Parente, G~Allan Johnson, Shivanand~P Lad, et~al.
\newblock A high-resolution interactive atlas of the human brainstem using magnetic resonance imaging.
\newblock \emph{Neuroimage}, 237:\penalty0 118135, 2021.

\bibitem[Babayan et~al.(2019)Babayan, Erbey, Kumral, Reinelt, Reiter, R{\"o}bbig, Schaare, Uhlig, Anwander, Bazin, et~al.]{babayan2019mind}
Anahit Babayan, Miray Erbey, Deniz Kumral, Janis~D Reinelt, Andrea~MF Reiter, Josefin R{\"o}bbig, H~Lina Schaare, Marie Uhlig, Alfred Anwander, Pierre-Louis Bazin, et~al.
\newblock A mind-brain-body dataset of mri, eeg, cognition, emotion, and peripheral physiology in young and old adults.
\newblock \emph{Scientific data}, 6\penalty0 (1):\penalty0 1--21, 2019.

\bibitem[Baid et~al.(2021)Baid, Ghodasara, Mohan, Bilello, Calabrese, Colak, Farahani, Kalpathy-Cramer, Kitamura, Pati, et~al.]{baid2021rsna}
Ujjwal Baid, Satyam Ghodasara, Suyash Mohan, Michel Bilello, Evan Calabrese, Errol Colak, Keyvan Farahani, Jayashree Kalpathy-Cramer, Felipe~C Kitamura, Sarthak Pati, et~al.
\newblock The rsna-asnr-miccai brats 2021 benchmark on brain tumor segmentation and radiogenomic classification.
\newblock \emph{arXiv preprint arXiv:2107.02314}, 2021.

\bibitem[Bannur et~al.(2024)Bannur, Bouzid, Castro, Schwaighofer, Bond-Taylor, Ilse, P{\'e}rez-Garc{\'\i}a, Salvatelli, Sharma, Meissen, et~al.]{bannur2024maira}
Shruthi Bannur, Kenza Bouzid, Daniel~C Castro, Anton Schwaighofer, Sam Bond-Taylor, Maximilian Ilse, Fernando P{\'e}rez-Garc{\'\i}a, Valentina Salvatelli, Harshita Sharma, Felix Meissen, et~al.
\newblock Maira-2: Grounded radiology report generation.
\newblock \emph{arXiv preprint arXiv:2406.04449}, 2024.

\bibitem[Billot et~al.(2023)Billot, Greve, Puonti, Thielscher, Van~Leemput, Fischl, Dalca, Iglesias, et~al.]{billot2023synthseg}
Benjamin Billot, Douglas~N Greve, Oula Puonti, Axel Thielscher, Koen Van~Leemput, Bruce Fischl, Adrian~V Dalca, Juan~Eugenio Iglesias, et~al.
\newblock Synthseg: Segmentation of brain mri scans of any contrast and resolution without retraining.
\newblock \emph{Medical image analysis}, 86:\penalty0 102789, 2023.

\bibitem[Boiko et~al.(2023)Boiko, MacKnight, and Gomes]{boiko2023emergent}
Daniil~A Boiko, Robert MacKnight, and Gabe Gomes.
\newblock Emergent autonomous scientific research capabilities of large language models.
\newblock \emph{arXiv preprint arXiv:2304.05332}, 2023.

\bibitem[Bran et~al.(2023)Bran, Cox, Schilter, Baldassari, White, and Schwaller]{bran2023chemcrow}
Andres~M Bran, Sam Cox, Oliver Schilter, Carlo Baldassari, Andrew~D White, and Philippe Schwaller.
\newblock Chemcrow: Augmenting large-language models with chemistry tools.
\newblock \emph{arXiv preprint arXiv:2304.05376}, 2023.

\bibitem[Butoi et~al.(2023)Butoi, Ortiz, Ma, Sabuncu, Guttag, and Dalca]{butoi2023universeg}
Victor~Ion Butoi, Jose Javier~Gonzalez Ortiz, Tianyu Ma, Mert~R Sabuncu, John Guttag, and Adrian~V Dalca.
\newblock Universeg: Universal medical image segmentation.
\newblock \emph{arXiv preprint arXiv:2304.06131}, 2023.

\bibitem[Chen et~al.(2023{\natexlab{a}})Chen, Hu, Wang, and Hong]{chen2023medblip}
Qiuhui Chen, Xinyue Hu, Zirui Wang, and Yi~Hong.
\newblock Medblip: Bootstrapping language-image pre-training from 3d medical images and texts.
\newblock \emph{arXiv preprint arXiv:2305.10799}, 2023{\natexlab{a}}.

\bibitem[Chen et~al.(2023{\natexlab{b}})Chen, Liu, Huang, Cheng, Arcucci, and Xiong]{chen2023generative}
Yinda Chen, Che Liu, Wei Huang, Sibo Cheng, Rossella Arcucci, and Zhiwei Xiong.
\newblock Generative text-guided 3d vision-language pretraining for unified medical image segmentation.
\newblock \emph{arXiv preprint arXiv:2306.04811}, 2023{\natexlab{b}}.

\bibitem[Cheng et~al.(2023)Cheng, Ye, Deng, Chen, Li, Wang, Su, Huang, Chen, Jiang, et~al.]{cheng2023sam}
Junlong Cheng, Jin Ye, Zhongying Deng, Jianpin Chen, Tianbin Li, Haoyu Wang, Yanzhou Su, Ziyan Huang, Jilong Chen, Lei Jiang, et~al.
\newblock Sam-med2d.
\newblock \emph{arXiv preprint arXiv:2308.16184}, 2023.

\bibitem[Czolbe \& Dalca(2023)Czolbe and Dalca]{czolbe2023neuralizer}
Steffen Czolbe and Adrian~V Dalca.
\newblock Neuralizer: General neuroimage analysis without re-training.
\newblock In \emph{Proceedings of the IEEE/CVF Conference on Computer Vision and Pattern Recognition}, pp.\  6217--6230, 2023.

\bibitem[Dey et~al.(2024)Dey, Billot, Wong, Wang, Ren, Grant, Dalca, and Golland]{dey2024learning}
Neel Dey, Benjamin Billot, Hallee~E Wong, Clinton~J Wang, Mengwei Ren, P~Ellen Grant, Adrian~V Dalca, and Polina Golland.
\newblock Learning general-purpose biomedical volume representations using randomized synthesis.
\newblock \emph{arXiv preprint arXiv:2411.02372}, 2024.

\bibitem[Elmahdy et~al.(2021)Elmahdy, Beljaards, Yousefi, Sokooti, Verbeek, Van Der~Heide, and Staring]{elmahdy2021joint}
Mohamed~S Elmahdy, Laurens Beljaards, Sahar Yousefi, Hessam Sokooti, Fons Verbeek, Uulke~A Van Der~Heide, and Marius Staring.
\newblock Joint registration and segmentation via multi-task learning for adaptive radiotherapy of prostate cancer.
\newblock \emph{IEEE Access}, 9:\penalty0 95551--95568, 2021.

\bibitem[Fischl(2012)]{fischl2012freesurfer}
Bruce Fischl.
\newblock Freesurfer.
\newblock \emph{Neuroimage}, 62\penalty0 (2):\penalty0 774--781, 2012.

\bibitem[Gopinath et~al.(2024)Gopinath, Hoopes, Alexander, Arnold, Balbastre, Billot, Casamitjana, Cheng, Chua, Edlow, et~al.]{gopinath2024synthetic}
Karthik Gopinath, Andrew Hoopes, Daniel~C Alexander, Steven~E Arnold, Yael Balbastre, Benjamin Billot, Adri{\`a} Casamitjana, You Cheng, Russ Yue~Zhi Chua, Brian~L Edlow, et~al.
\newblock Synthetic data in generalizable, learning-based neuroimaging.
\newblock \emph{Imaging Neuroscience}, 2:\penalty0 1--22, 2024.

\bibitem[Gou et~al.(2023)Gou, Shao, Gong, Yang, Huang, Duan, Chen, et~al.]{gou2023tora}
Zhibin Gou, Zhihong Shao, Yeyun Gong, Yujiu Yang, Minlie Huang, Nan Duan, Weizhu Chen, et~al.
\newblock Tora: A tool-integrated reasoning agent for mathematical problem solving.
\newblock \emph{arXiv preprint arXiv:2309.17452}, 2023.

\bibitem[Graham et~al.(2023)Graham, Vu, Jahanifar, Raza, Minhas, Snead, and Rajpoot]{graham2023one}
Simon Graham, Quoc~Dang Vu, Mostafa Jahanifar, Shan E~Ahmed Raza, Fayyaz Minhas, David Snead, and Nasir Rajpoot.
\newblock One model is all you need: multi-task learning enables simultaneous histology image segmentation and classification.
\newblock \emph{Medical Image Analysis}, 83:\penalty0 102685, 2023.

\bibitem[Greve et~al.(2021)Greve, Billot, Cordero, Hoopes, Hoffmann, Dalca, Fischl, Iglesias, and Augustinack]{greve2021deep}
Douglas~N Greve, Benjamin Billot, Devani Cordero, Andrew Hoopes, Malte Hoffmann, Adrian~V Dalca, Bruce Fischl, Juan~Eugenio Iglesias, and Jean~C Augustinack.
\newblock A deep learning toolbox for automatic segmentation of subcortical limbic structures from mri images.
\newblock \emph{Neuroimage}, 244:\penalty0 118610, 2021.

\bibitem[Gupta \& Kembhavi(2023)Gupta and Kembhavi]{gupta2023visual}
Tanmay Gupta and Aniruddha Kembhavi.
\newblock Visual programming: Compositional visual reasoning without training.
\newblock In \emph{Proceedings of the IEEE/CVF Conference on Computer Vision and Pattern Recognition}, pp.\  14953--14962, 2023.

\bibitem[Hanke et~al.(2014)Hanke, Baumgartner, Ibe, Kaule, Pollmann, Speck, Zinke, and Stadler]{hanke2014high}
Michael Hanke, Florian~J Baumgartner, Pierre Ibe, Falko~R Kaule, Stefan Pollmann, Oliver Speck, Wolf Zinke, and J{\"o}rg Stadler.
\newblock A high-resolution 7-tesla fmri dataset from complex natural stimulation with an audio movie.
\newblock \emph{Scientific data}, 1\penalty0 (1):\penalty0 1--18, 2014.

\bibitem[Henschel et~al.(2020)Henschel, Conjeti, Estrada, Diers, Fischl, and Reuter]{henschel2020fastsurfer}
Leonie Henschel, Sailesh Conjeti, Santiago Estrada, Kersten Diers, Bruce Fischl, and Martin Reuter.
\newblock Fastsurfer - a fast and accurate deep learning based neuroimaging pipeline.
\newblock \emph{NeuroImage}, 219:\penalty0 117012, 2020.

\bibitem[Hernandez~Petzsche et~al.(2022)Hernandez~Petzsche, de~la Rosa, Hanning, Wiest, Valenzuela, Reyes, Meyer, Liew, Kofler, Ezhov, et~al.]{hernandez2022isles}
Moritz~R Hernandez~Petzsche, Ezequiel de~la Rosa, Uta Hanning, Roland Wiest, Waldo Valenzuela, Mauricio Reyes, Maria Meyer, Sook-Lei Liew, Florian Kofler, Ivan Ezhov, et~al.
\newblock Isles 2022: A multi-center magnetic resonance imaging stroke lesion segmentation dataset.
\newblock \emph{Scientific data}, 9\penalty0 (1):\penalty0 762, 2022.

\bibitem[Hilbert et~al.(2020)Hilbert, Madai, Akay, Aydin, Behland, Sobesky, Galinovic, Khalil, Taha, Wuerfel, et~al.]{hilbert2020brave}
Adam Hilbert, Vince~I Madai, Ela~M Akay, Orhun~U Aydin, Jonas Behland, Jan Sobesky, Ivana Galinovic, Ahmed~A Khalil, Abdel~A Taha, Jens Wuerfel, et~al.
\newblock Brave-net: fully automated arterial brain vessel segmentation in patients with cerebrovascular disease.
\newblock \emph{Frontiers in artificial intelligence}, 3:\penalty0 552258, 2020.

\bibitem[Hoffmann et~al.(2024)Hoffmann, Hoopes, Greve, Fischl, and Dalca]{hoffmann2024}
Malte Hoffmann, Andrew Hoopes, Douglas~N. Greve, Bruce Fischl, and Adrian~V. Dalca.
\newblock {Anatomy-aware and acquisition-agnostic joint registration with SynthMorph}.
\newblock \emph{Imaging Neuroscience}, 2:\penalty0 1--33, 06 2024.
\newblock ISSN 2837-6056.
\newblock \doi{10.1162/imag_a_00197}.

\bibitem[Hoopes et~al.(2022)Hoopes, Mora, Dalca, Fischl, and Hoffmann]{hoopes2022synthstrip}
Andrew Hoopes, Jocelyn~S Mora, Adrian~V Dalca, Bruce Fischl, and Malte Hoffmann.
\newblock Synthstrip: Skull-stripping for any brain image.
\newblock \emph{NeuroImage}, 260:\penalty0 119474, 2022.

\bibitem[Hssayeni et~al.(2020)Hssayeni, Croock, Salman, Al-Khafaji, Yahya, and Ghoraani]{hssayeni2020intracranial}
Murtadha~D Hssayeni, Muayad~S Croock, Aymen~D Salman, Hassan~Falah Al-Khafaji, Zakaria~A Yahya, and Behnaz Ghoraani.
\newblock Intracranial hemorrhage segmentation using a deep convolutional model.
\newblock \emph{Data}, 5\penalty0 (1):\penalty0 14, 2020.

\bibitem[Huang et~al.(2022)Huang, Xia, Xiao, Chan, Liang, Florence, Zeng, Tompson, Mordatch, Chebotar, et~al.]{huang2022inner}
Wenlong Huang, Fei Xia, Ted Xiao, Harris Chan, Jacky Liang, Pete Florence, Andy Zeng, Jonathan Tompson, Igor Mordatch, Yevgen Chebotar, et~al.
\newblock Inner monologue: Embodied reasoning through planning with language models.
\newblock \emph{arXiv preprint arXiv:2207.05608}, 2022.

\bibitem[Isensee et~al.(2025)Isensee, Rokuss, Kr{\"a}mer, Dinkelacker, Ravindran, Stritzke, Hamm, Wald, Langenberg, Ulrich, et~al.]{isensee2025nninteractive}
Fabian Isensee, Maximilian Rokuss, Lars Kr{\"a}mer, Stefan Dinkelacker, Ashis Ravindran, Florian Stritzke, Benjamin Hamm, Tassilo Wald, Moritz Langenberg, Constantin Ulrich, et~al.
\newblock nninteractive: Redefining 3d promptable segmentation.
\newblock \emph{arXiv preprint arXiv:2503.08373}, 2025.

\bibitem[Jenkinson et~al.(2012)Jenkinson, Beckmann, Behrens, Woolrich, and Smith]{jenkinson2012fsl}
Mark Jenkinson, Christian~F Beckmann, Timothy~EJ Behrens, Mark~W Woolrich, and Stephen~M Smith.
\newblock Fsl.
\newblock \emph{Neuroimage}, 62\penalty0 (2):\penalty0 782--790, 2012.

\bibitem[Johnson et~al.(2019)Johnson, Pollard, Berkowitz, Greenbaum, Lungren, Deng, Mark, and Horng]{johnson2019mimic}
Alistair~EW Johnson, Tom~J Pollard, Seth~J Berkowitz, Nathaniel~R Greenbaum, Matthew~P Lungren, Chih-ying Deng, Roger~G Mark, and Steven Horng.
\newblock Mimic-cxr, a de-identified publicly available database of chest radiographs with free-text reports.
\newblock \emph{Scientific data}, 6\penalty0 (1):\penalty0 317, 2019.

\bibitem[Ke et~al.(2025)Ke, Hsu, Cai, Ma, Zheng, Wu, Huang, Wang, Haghighi, Haffari, et~al.]{ke2025explain}
Fucai Ke, Joy Hsu, Zhixi Cai, Zixian Ma, Xin Zheng, Xindi Wu, Sukai Huang, Weiqing Wang, Pari~Delir Haghighi, Gholamreza Haffari, et~al.
\newblock Explain before you answer: A survey on compositional visual reasoning.
\newblock \emph{arXiv preprint arXiv:2508.17298}, 2025.

\bibitem[Kingma \& Ba(2014)Kingma and Ba]{kingma2014adam}
Diederik~P Kingma and Jimmy Ba.
\newblock Adam: A method for stochastic optimization.
\newblock \emph{arXiv preprint arXiv:1412.6980}, 2014.

\bibitem[Kuijf et~al.(2019)Kuijf, Biesbroek, De~Bresser, Heinen, Andermatt, Bento, Berseth, Belyaev, Cardoso, Casamitjana, et~al.]{kuijf2019standardized}
Hugo~J Kuijf, J~Matthijs Biesbroek, Jeroen De~Bresser, Rutger Heinen, Simon Andermatt, Mariana Bento, Matt Berseth, Mikhail Belyaev, M~Jorge Cardoso, Adria Casamitjana, et~al.
\newblock Standardized assessment of automatic segmentation of white matter hyperintensities and results of the wmh segmentation challenge.
\newblock \emph{IEEE transactions on medical imaging}, 38\penalty0 (11):\penalty0 2556--2568, 2019.

\bibitem[LaBella et~al.(2024)LaBella, Khanna, McBurney-Lin, Mclean, Nedelec, Rashid, Tahon, Altes, Baid, Bhalerao, et~al.]{labella2024multi}
Dominic LaBella, Omaditya Khanna, Shan McBurney-Lin, Ryan Mclean, Pierre Nedelec, Arif~S Rashid, Nourel~Hoda Tahon, Talissa Altes, Ujjwal Baid, Radhika Bhalerao, et~al.
\newblock A multi-institutional meningioma mri dataset for automated multi-sequence image segmentation.
\newblock \emph{Scientific data}, 11\penalty0 (1):\penalty0 496, 2024.

\bibitem[LaMontagne et~al.(2019)LaMontagne, Benzinger, Morris, Keefe, Hornbeck, Xiong, Grant, Hassenstab, Moulder, Vlassenko, et~al.]{lamontagne2019oasis}
Pamela~J LaMontagne, Tammie~LS Benzinger, John~C Morris, Sarah Keefe, Russ Hornbeck, Chengjie Xiong, Elizabeth Grant, Jason Hassenstab, Krista Moulder, Andrei~G Vlassenko, et~al.
\newblock Oasis-3: longitudinal neuroimaging, clinical, and cognitive dataset for normal aging and alzheimer disease.
\newblock \emph{MedRxiv}, pp.\  2019--12, 2019.

\bibitem[Li et~al.(2024)Li, Yan, Pan, Xu, Luo, Ji, Liu, Dong, Lin, and Wang]{li2024mmedagent}
Binxu Li, Tiankai Yan, Yuanting Pan, Zhe Xu, Jie Luo, Ruiyang Ji, Shilong Liu, Haoyu Dong, Zihao Lin, and Yixin Wang.
\newblock Mmedagent: Learning to use medical tools with multi-modal agent.
\newblock \emph{arXiv preprint arXiv:2407.02483}, 2024.

\bibitem[Liew et~al.(2022)Liew, Lo, Donnelly, Zavaliangos-Petropulu, Jeong, Barisano, Hutton, Simon, Juliano, Suri, et~al.]{liew2022large}
Sook-Lei Liew, Bethany~P Lo, Miranda~R Donnelly, Artemis Zavaliangos-Petropulu, Jessica~N Jeong, Giuseppe Barisano, Alexandre Hutton, Julia~P Simon, Julia~M Juliano, Anisha Suri, et~al.
\newblock A large, curated, open-source stroke neuroimaging dataset to improve lesion segmentation algorithms.
\newblock \emph{Scientific data}, 9\penalty0 (1):\penalty0 320, 2022.

\bibitem[Lin et~al.(2023)Lin, Zhao, Zhang, Wu, Zhang, Wang, and Xie]{lin2023pmc}
Weixiong Lin, Ziheng Zhao, Xiaoman Zhang, Chaoyi Wu, Ya~Zhang, Yanfeng Wang, and Weidi Xie.
\newblock Pmc-clip: Contrastive language-image pre-training using biomedical documents.
\newblock In \emph{International Conference on Medical Image Computing and Computer-Assisted Intervention}, pp.\  525--536. Springer, 2023.

\bibitem[Liu et~al.(2023)Liu, Ouyang, Chen, Quilodr{\'a}n-Casas, Ma, Fu, Guo, Shah, Bai, and Arcucci]{liu2023t3d}
Che Liu, Cheng Ouyang, Yinda Chen, Cesar~C{\'e}sar Quilodr{\'a}n-Casas, Lei Ma, Jie Fu, Yike Guo, Anand Shah, Wenjia Bai, and Rossella Arcucci.
\newblock T3d: Towards 3d medical image understanding through vision-language pre-training.
\newblock \emph{arXiv preprint arXiv:2312.01529}, 2023.

\bibitem[Liu et~al.(2021)Liu, Hsu, Xu, Ramachandran, Wang, Miller, Hillis, Faria, K, Wintermark, Warach, Albers, Davis, Grotta, Hacke, Kang, Kidwell, Koroshetz, Lees, Lev, Liebeskind, Sorensen, Thijs, Thomalla, Wardlaw, and Luby]{liu2021deep}
Chin-Fu Liu, Johnny T.~C. Hsu, Xin Xu, Sandhya Ramachandran, Victor Wang, Michael~I. Miller, Argye~Elizabeth Hillis, Andreia~Vasconcellos Faria, Max Steven J. Gregory W. Stephen M. James C. Werner Do Wintermark Warach Albers Davis Grotta Hacke~Kang K, Max Wintermark, Steven~J. Warach, Gregory~W. Albers, Stephen~M. Davis, James~Charles Grotta, Werner Hacke, Dong-Wha Kang, Chelsea Kidwell, Walter~J. Koroshetz, Kennedy~R. Lees, Michael~H. Lev, David~S. Liebeskind, A.~Gregory Sorensen, Vincent~N. Thijs, G{\"o}tz Thomalla, Joanna~Marguerite Wardlaw, and Marie Luby.
\newblock Deep learning-based detection and segmentation of diffusion abnormalities in acute ischemic stroke.
\newblock \emph{Communications Medicine}, 1, 2021.

\bibitem[Liu et~al.(2025)Liu, Puonti, Hu, Gopinath, Sorby-Adams, Alexander, Kimberly, and Iglesias]{liu2025modality}
Peirong Liu, Oula Puonti, Xiaoling Hu, Karthik Gopinath, Annabel Sorby-Adams, Daniel~C Alexander, W~Taylor Kimberly, and Juan~E Iglesias.
\newblock A modality-agnostic multi-task foundation model for human brain imaging.
\newblock \emph{arXiv preprint arXiv:2509.00549}, 2025.

\bibitem[Livne et~al.(2019)Livne, Rieger, Aydin, Taha, Akay, Kossen, Sobesky, Kelleher, Hildebrand, Frey, et~al.]{livne2019u}
Michelle Livne, Jana Rieger, Orhun~Utku Aydin, Abdel~Aziz Taha, Ela~Marie Akay, Tabea Kossen, Jan Sobesky, John~D Kelleher, Kristian Hildebrand, Dietmar Frey, et~al.
\newblock A u-net deep learning framework for high performance vessel segmentation in patients with cerebrovascular disease.
\newblock \emph{Frontiers in neuroscience}, 13:\penalty0 97, 2019.

\bibitem[Luo et~al.(2021)Luo, Wang, Song, Zhang, Aertsen, Deprest, Ourselin, Vercauteren, and Zhang]{luo2021mideepseg}
Xiangde Luo, Guotai Wang, Tao Song, Jingyang Zhang, Michael Aertsen, Jan Deprest, Sebastien Ourselin, Tom Vercauteren, and Shaoting Zhang.
\newblock Mideepseg: Minimally interactive segmentation of unseen objects from medical images using deep learning.
\newblock \emph{Medical image analysis}, 72:\penalty0 102102, 2021.

\bibitem[Ma et~al.(2024)Ma, He, Li, Han, You, and Wang]{ma2024segment}
Jun Ma, Yuting He, Feifei Li, Lin Han, Chenyu You, and Bo~Wang.
\newblock Segment anything in medical images.
\newblock \emph{Nature Communications}, 15\penalty0 (1):\penalty0 654, 2024.

\bibitem[Marcus et~al.(2007)Marcus, Wang, Parker, Csernansky, Morris, and Buckner]{marcus2007open}
Daniel~S Marcus, Tracy~H Wang, Jamie Parker, John~G Csernansky, John~C Morris, and Randy~L Buckner.
\newblock Open access series of imaging studies (oasis): cross-sectional mri data in young, middle aged, nondemented, and demented older adults.
\newblock \emph{Journal of cognitive neuroscience}, 19\penalty0 (9):\penalty0 1498--1507, 2007.

\bibitem[M{\'e}rida et~al.(2021)M{\'e}rida, Jung, Bouvard, Le~Bars, Lancelot, Lavenne, Bouillot, Redout{\'e}, Hammers, and Costes]{merida2021cermep}
In{\'e}s M{\'e}rida, Julien Jung, Sandrine Bouvard, Didier Le~Bars, Sophie Lancelot, Franck Lavenne, Caroline Bouillot, J{\'e}r{\^o}me Redout{\'e}, Alexander Hammers, and Nicolas Costes.
\newblock Cermep-idb-mrxfdg: A database of 37 normal adult human brain [18f] fdg pet, t1 and flair mri, and ct images available for research.
\newblock \emph{EJNMMI research}, 11\penalty0 (1):\penalty0 1--10, 2021.

\bibitem[Min et~al.(2021)Min, Lewis, Zettlemoyer, and Hajishirzi]{min2021metaicl}
Sewon Min, Mike Lewis, Luke Zettlemoyer, and Hannaneh Hajishirzi.
\newblock Metaicl: Learning to learn in context.
\newblock \emph{arXiv preprint arXiv:2110.15943}, 2021.

\bibitem[Ouyang et~al.(2022)Ouyang, Biffi, Chen, Kart, Qiu, and Rueckert]{ouyang2022self}
Cheng Ouyang, Carlo Biffi, Chen Chen, Turkay Kart, Huaqi Qiu, and Daniel Rueckert.
\newblock Self-supervised learning for few-shot medical image segmentation.
\newblock \emph{IEEE Transactions on Medical Imaging}, 41\penalty0 (7):\penalty0 1837--1848, 2022.

\bibitem[Paszke et~al.(2019)Paszke, Gross, Massa, Lerer, Bradbury, Chanan, Killeen, Lin, Gimelshein, Antiga, et~al.]{paszke2019pytorch}
Adam Paszke, Sam Gross, Francisco Massa, Adam Lerer, James Bradbury, Gregory Chanan, Trevor Killeen, Zeming Lin, Natalia Gimelshein, Luca Antiga, et~al.
\newblock Pytorch: An imperative style, high-performance deep learning library.
\newblock \emph{Advances in neural information processing systems}, 32, 2019.

\bibitem[Pauli et~al.(2018)Pauli, Nili, and Tyszka]{pauli2018high}
Wolfgang~M Pauli, Amanda~N Nili, and J~Michael Tyszka.
\newblock A high-resolution probabilistic in vivo atlas of human subcortical brain nuclei.
\newblock \emph{Scientific data}, 5\penalty0 (1):\penalty0 1--13, 2018.

\bibitem[P{\'e}rez-Garc{\'\i}a et~al.(2020)P{\'e}rez-Garc{\'\i}a, Rodionov, Alim-Marvasti, Sparks, Duncan, and Ourselin]{perez2020simulation}
Fernando P{\'e}rez-Garc{\'\i}a, Roman Rodionov, Ali Alim-Marvasti, Rachel Sparks, John~S Duncan, and S{\'e}bastien Ourselin.
\newblock Simulation of brain resection for cavity segmentation using self-supervised and semi-supervised learning.
\newblock In \emph{International Conference on Medical Image Computing and Computer-Assisted Intervention}, pp.\  115--125. Springer, 2020.

\bibitem[Pinho et~al.(2018)Pinho, Amadon, Ruest, Fabre, Dohmatob, Denghien, Ginisty, Becuwe-Desmidt, Roger, Laurier, et~al.]{pinho2018individual}
Ana~Lu{\'\i}sa Pinho, Alexis Amadon, Torsten Ruest, Murielle Fabre, Elvis Dohmatob, Isabelle Denghien, Chantal Ginisty, S{\'e}verine Becuwe-Desmidt, S{\'e}verine Roger, Laurence Laurier, et~al.
\newblock Individual brain charting, a high-resolution fmri dataset for cognitive mapping.
\newblock \emph{Scientific data}, 5\penalty0 (1):\penalty0 1--15, 2018.

\bibitem[Popa et~al.(2025)Popa, Vișa, and Șofariu]{popa2025pedims}
Maria Popa, Gabriela~Adriana Vișa, and Ciprian~Radu Șofariu.
\newblock Pedims: A pediatric multiple sclerosis lesion segmentation dataset.
\newblock \emph{Scientific Data}, 12\penalty0 (1):\penalty0 1184, 2025.

\bibitem[Rakic et~al.(2024)Rakic, Wong, Ortiz, Cimini, Guttag, and Dalca]{rakic2024tyche}
Marianne Rakic, Hallee~E Wong, Jose Javier~Gonzalez Ortiz, Beth Cimini, John Guttag, and Adrian~V Dalca.
\newblock Tyche: Stochastic in-context learning for medical image segmentation.
\newblock \emph{arXiv preprint arXiv:2401.13650}, 2024.

\bibitem[Rana et~al.(2023)Rana, Haviland, Garg, Abou-Chakra, Reid, and Suenderhauf]{rana2023sayplan}
Krishan Rana, Jesse Haviland, Sourav Garg, Jad Abou-Chakra, Ian Reid, and Niko Suenderhauf.
\newblock Sayplan: Grounding large language models using 3d scene graphs for scalable task planning.
\newblock \emph{arXiv preprint arXiv:2307.06135}, 2023.

\bibitem[Reuter et~al.(2012)Reuter, Schmansky, Rosas, and Fischl]{reuter2012within}
Martin Reuter, Nicholas~J Schmansky, H~Diana Rosas, and Bruce Fischl.
\newblock Within-subject template estimation for unbiased longitudinal image analysis.
\newblock \emph{Neuroimage}, 61\penalty0 (4):\penalty0 1402--1418, 2012.

\bibitem[Ronneberger et~al.(2015)Ronneberger, Fischer, and Brox]{ronneberger2015u}
Olaf Ronneberger, Philipp Fischer, and Thomas Brox.
\newblock U-net: Convolutional networks for biomedical image segmentation.
\newblock In \emph{Medical image computing and computer-assisted intervention--MICCAI 2015: 18th international conference, Munich, Germany, October 5-9, 2015, proceedings, part III 18}, pp.\  234--241. Springer, 2015.

\bibitem[Roy et~al.(2020)Roy, Siddiqui, P{\"o}lsterl, Navab, and Wachinger]{roy2020squeeze}
Abhijit~Guha Roy, Shayan Siddiqui, Sebastian P{\"o}lsterl, Nassir Navab, and Christian Wachinger.
\newblock ‘squeeze \& excite’guided few-shot segmentation of volumetric images.
\newblock \emph{Medical image analysis}, 59:\penalty0 101587, 2020.

\bibitem[Ruan et~al.(2023)Ruan, Chen, Zhang, Xu, Bao, Mao, Li, Zeng, Zhao, et~al.]{ruan2023tptu}
Jingqing Ruan, Yihong Chen, Bin Zhang, Zhiwei Xu, Tianpeng Bao, Hangyu Mao, Ziyue Li, Xingyu Zeng, Rui Zhao, et~al.
\newblock Tptu: Task planning and tool usage of large language model-based ai agents.
\newblock In \emph{NeurIPS 2023 Foundation Models for Decision Making Workshop}, 2023.

\bibitem[Subramanian et~al.(2023)Subramanian, Narasimhan, Khangaonkar, Yang, Nagrani, Schmid, Zeng, Darrell, and Klein]{subramanian2023modular}
Sanjay Subramanian, Medhini Narasimhan, Kushal Khangaonkar, Kevin Yang, Arsha Nagrani, Cordelia Schmid, Andy Zeng, Trevor Darrell, and Dan Klein.
\newblock Modular visual question answering via code generation.
\newblock \emph{arXiv preprint arXiv:2306.05392}, 2023.

\bibitem[Sur{\'\i}s et~al.(2023)Sur{\'\i}s, Menon, and Vondrick]{suris2023vipergpt}
D{\'\i}dac Sur{\'\i}s, Sachit Menon, and Carl Vondrick.
\newblock Vipergpt: Visual inference via python execution for reasoning.
\newblock In \emph{Proceedings of the IEEE/CVF International Conference on Computer Vision}, pp.\  11888--11898, 2023.

\bibitem[Tellez et~al.(2020)Tellez, H{\"o}ppener, Verhoef, Gr{\"u}nhagen, Nierop, Drozdzal, Laak, and Ciompi]{tellez2020extending}
David Tellez, Diederik H{\"o}ppener, Cornelis Verhoef, Dirk Gr{\"u}nhagen, Pieter Nierop, Michal Drozdzal, Jeroen Laak, and Francesco Ciompi.
\newblock Extending unsupervised neural image compression with supervised multitask learning.
\newblock In \emph{Medical Imaging with Deep Learning}, pp.\  770--783. PMLR, 2020.

\bibitem[Tobin et~al.(2017)Tobin, Fong, Ray, Schneider, Zaremba, and Abbeel]{tobin2017domain}
Josh Tobin, Rachel Fong, Alex Ray, Jonas Schneider, Wojciech Zaremba, and Pieter Abbeel.
\newblock Domain randomization for transferring deep neural networks from simulation to the real world.
\newblock In \emph{2017 IEEE/RSJ international conference on intelligent robots and systems (IROS)}, pp.\  23--30. IEEE, 2017.

\bibitem[Touvron et~al.(2023)Touvron, Martin, Stone, Albert, Almahairi, Babaei, Bashlykov, Batra, Bhargava, Bhosale, et~al.]{touvron2023llama}
Hugo Touvron, Louis Martin, Kevin Stone, Peter Albert, Amjad Almahairi, Yasmine Babaei, Nikolay Bashlykov, Soumya Batra, Prajjwal Bhargava, Shruti Bhosale, et~al.
\newblock Llama 2: Open foundation and fine-tuned chat models.
\newblock \emph{arXiv preprint arXiv:2307.09288}, 2023.

\bibitem[Wang et~al.(2023{\natexlab{a}})Wang, Xie, Jiang, Mandlekar, Xiao, Zhu, Fan, and Anandkumar]{wang2023voyager}
Guanzhi Wang, Yuqi Xie, Yunfan Jiang, Ajay Mandlekar, Chaowei Xiao, Yuke Zhu, Linxi Fan, and Anima Anandkumar.
\newblock Voyager: An open-ended embodied agent with large language models.
\newblock \emph{arXiv preprint arXiv:2305.16291}, 2023{\natexlab{a}}.

\bibitem[Wang et~al.(2023{\natexlab{b}})Wang, Liu, Wang, and Zhou]{wang2023metransformer}
Zhanyu Wang, Lingqiao Liu, Lei Wang, and Luping Zhou.
\newblock Metransformer: Radiology report generation by transformer with multiple learnable expert tokens.
\newblock In \emph{Proceedings of the IEEE/CVF Conference on Computer Vision and Pattern Recognition}, pp.\  11558--11567, 2023{\natexlab{b}}.

\bibitem[Wang et~al.(2023{\natexlab{c}})Wang, Liu, Wang, and Zhou]{wang2023r2gengpt}
Zhanyu Wang, Lingqiao Liu, Lei Wang, and Luping Zhou.
\newblock R2gengpt: Radiology report generation with frozen llms.
\newblock \emph{Meta-Radiology}, 1\penalty0 (3):\penalty0 100033, 2023{\natexlab{c}}.

\bibitem[Wang et~al.(2023{\natexlab{d}})Wang, Cai, Chen, Liu, Ma, and Liang]{wang2023describe}
Zihao Wang, Shaofei Cai, Guanzhou Chen, Anji Liu, Xiaojian Ma, and Yitao Liang.
\newblock Describe, explain, plan and select: Interactive planning with large language models enables open-world multi-task agents.
\newblock \emph{arXiv preprint arXiv:2302.01560}, 2023{\natexlab{d}}.

\bibitem[Weber et~al.(2021)Weber, Carrillo, Jagust, Jack~Jr, Shaw, Trojanowski, Saykin, Beckett, Sur, Rao, et~al.]{weber2021worldwide}
Christopher~J Weber, Maria~C Carrillo, William Jagust, Clifford~R Jack~Jr, Leslie~M Shaw, John~Q Trojanowski, Andrew~J Saykin, Laurel~A Beckett, Cyrille Sur, Naren~P Rao, et~al.
\newblock The worldwide alzheimer's disease neuroimaging initiative: Adni-3 updates and global perspectives.
\newblock \emph{Alzheimer's \& Dementia: Translational Research \& Clinical Interventions}, 7\penalty0 (1):\penalty0 e12226, 2021.

\bibitem[Wolf et~al.(2019)Wolf, Debut, Sanh, Chaumond, Delangue, Moi, Cistac, Rault, Louf, Funtowicz, et~al.]{wolf2019huggingface}
Thomas Wolf, Lysandre Debut, Victor Sanh, Julien Chaumond, Clement Delangue, Anthony Moi, Pierric Cistac, Tim Rault, R{\'e}mi Louf, Morgan Funtowicz, et~al.
\newblock Huggingface's transformers: State-of-the-art natural language processing.
\newblock \emph{arXiv preprint arXiv:1910.03771}, 2019.

\bibitem[Wong et~al.(2023)Wong, Rakic, Guttag, and Dalca]{wong2023scribbleprompt}
Hallee~E Wong, Marianne Rakic, John Guttag, and Adrian~V Dalca.
\newblock Scribbleprompt: Fast and flexible interactive segmentation for any medical image.
\newblock \emph{arXiv preprint arXiv:2312.07381}, 2023.

\bibitem[Wu et~al.(2023)Wu, Xie, Zhang, Ge, Yaxley, Bahadir, Wu, Liu, and To]{wu2023bhsd}
Biao Wu, Yutong Xie, Zeyu Zhang, Jinchao Ge, Kaspar Yaxley, Suzan Bahadir, Qi~Wu, Yifan Liu, and Minh-Son To.
\newblock Bhsd: A 3d multi-class brain hemorrhage segmentation dataset.
\newblock In \emph{International workshop on machine learning in medical imaging}, pp.\  147--156. Springer, 2023.

\bibitem[Wu et~al.(2025)Wu, Zhang, Zhang, Hui, Wang, and Xie]{wu2025towards}
Chaoyi Wu, Xiaoman Zhang, Ya~Zhang, Hui Hui, Yanfeng Wang, and Weidi Xie.
\newblock Towards generalist foundation model for radiology by leveraging web-scale 2d\&3d medical data.
\newblock \emph{Nature Communications}, 16\penalty0 (1):\penalty0 7866, 2025.

\bibitem[Xie et~al.(2021)Xie, Raghunathan, Liang, and Ma]{xie2021explanation}
Sang~Michael Xie, Aditi Raghunathan, Percy Liang, and Tengyu Ma.
\newblock An explanation of in-context learning as implicit bayesian inference.
\newblock \emph{arXiv preprint arXiv:2111.02080}, 2021.

\bibitem[Yang et~al.(2023)Yang, Li, Wang, Lin, Azarnasab, Ahmed, Liu, Liu, Zeng, and Wang]{yang2023mm}
Zhengyuan Yang, Linjie Li, Jianfeng Wang, Kevin Lin, Ehsan Azarnasab, Faisal Ahmed, Zicheng Liu, Ce~Liu, Michael Zeng, and Lijuan Wang.
\newblock Mm-react: Prompting chatgpt for multimodal reasoning and action.
\newblock \emph{arXiv preprint arXiv:2303.11381}, 2023.

\bibitem[Yao et~al.(2022)Yao, Zhao, Yu, Du, Shafran, Narasimhan, and Cao]{yao2022react}
Shunyu Yao, Jeffrey Zhao, Dian Yu, Nan Du, Izhak Shafran, Karthik Narasimhan, and Yuan Cao.
\newblock React: Synergizing reasoning and acting in language models.
\newblock \emph{arXiv preprint arXiv:2210.03629}, 2022.

\bibitem[Zhang et~al.(2023{\natexlab{a}})Zhang, Xu, Usuyama, Xu, Bagga, Tinn, Preston, Rao, Wei, Valluri, et~al.]{zhang2023biomedclip}
Sheng Zhang, Yanbo Xu, Naoto Usuyama, Hanwen Xu, Jaspreet Bagga, Robert Tinn, Sam Preston, Rajesh Rao, Mu~Wei, Naveen Valluri, et~al.
\newblock Biomedclip: a multimodal biomedical foundation model pretrained from fifteen million scientific image-text pairs.
\newblock \emph{arXiv preprint arXiv:2303.00915}, 2023{\natexlab{a}}.

\bibitem[Zhang et~al.(2023{\natexlab{b}})Zhang, Wu, Zhao, Lin, Zhang, Wang, and Xie]{zhang2023pmc}
Xiaoman Zhang, Chaoyi Wu, Ziheng Zhao, Weixiong Lin, Ya~Zhang, Yanfeng Wang, and Weidi Xie.
\newblock Pmc-vqa: Visual instruction tuning for medical visual question answering.
\newblock \emph{arXiv preprint arXiv:2305.10415}, 2023{\natexlab{b}}.

\bibitem[Zhang et~al.(2024)Zhang, Ou, Basaran, Visentin, Qiao, Gu, Ouyang, Liu, Matthews, Ye, et~al.]{zhang2024foundation}
Xinru Zhang, Ni~Ou, Berke~Doga Basaran, Marco Visentin, Mengyun Qiao, Renyang Gu, Cheng Ouyang, Yaou Liu, Paul~M Matthews, Chuyang Ye, et~al.
\newblock A foundation model for brain lesion segmentation with mixture of modality experts.
\newblock In \emph{International Conference on Medical Image Computing and Computer-Assisted Intervention}, pp.\  379--389. Springer, 2024.

\bibitem[Zhao et~al.(2024)Zhao, Gu, Yang, Usuyama, Lee, Naumann, Gao, Crabtree, Piening, Bifulco, et~al.]{zhao2024biomedparse}
Theodore Zhao, Yu~Gu, Jianwei Yang, Naoto Usuyama, Ho~Hin Lee, Tristan Naumann, Jianfeng Gao, Angela Crabtree, Brian Piening, Carlo Bifulco, et~al.
\newblock Biomedparse: a biomedical foundation model for image parsing of everything everywhere all at once.
\newblock \emph{arXiv preprint arXiv:2405.12971}, 2024.

\bibitem[Zhao et~al.(2025{\natexlab{a}})Zhao, Lee, Santamaria-Pang, Codella, Kiblawi, Gu, Fang, Teng, Sangani, Tarapov, Lungren, Blondeel, Naumann, Usuyama, Wang, Vozila, Poon, and Wei]{zhao2025biomedparsev}
Theodore Zhao, Ho~Hin Lee, Alberto Santamaria-Pang, Noel~C Codella, Sid Kiblawi, Yu~Gu, Yu~Fang, Wenxuan Teng, Naiteek Sangani, Ivan Tarapov, Matthew~P. Lungren, Matthias Blondeel, Tristan Naumann, Naoto Usuyama, Sheng Wang, Paul Vozila, Hoifung Poon, and Mu~Wei.
\newblock Biomedparse-v : Scaling foundation model for universal text-guided volumetric biomedical image segmentation.
\newblock In \emph{Foundation Models for 3D Biomedical Image Segmentation Workshop at CVPR}, 2025{\natexlab{a}}.

\bibitem[Zhao et~al.(2025{\natexlab{b}})Zhao, Zhang, Wu, Zhang, Zhou, Zhang, Wang, and Xie]{zhao2025large}
Ziheng Zhao, Yao Zhang, Chaoyi Wu, Xiaoman Zhang, Xiao Zhou, Ya~Zhang, Yanfeng Wang, and Weidi Xie.
\newblock Large-vocabulary segmentation for medical images with text prompts.
\newblock \emph{npj Digital Medicine}, 8\penalty0 (1):\penalty0 566, 2025{\natexlab{b}}.

\bibitem[Zhou et~al.(2024)Zhou, Adithan, Acosta, Topol, and Rajpurkar]{zhou2024generalist}
Hong-Yu Zhou, Subathra Adithan, Juli{\'a}n~Nicol{\'a}s Acosta, Eric~J Topol, and Pranav Rajpurkar.
\newblock A generalist learner for multifaceted medical image interpretation.
\newblock \emph{arXiv preprint arXiv:2405.07988}, 2024.

\bibitem[Zhu et~al.(2023)Zhu, Chen, Tian, Tao, Su, Yang, Huang, Li, Lu, Wang, et~al.]{zhu2023ghost}
Xizhou Zhu, Yuntao Chen, Hao Tian, Chenxin Tao, Weijie Su, Chenyu Yang, Gao Huang, Bin Li, Lewei Lu, Xiaogang Wang, et~al.
\newblock Ghost in the minecraft: Generally capable agents for open-world environments via large language models with text-based knowledge and memory.
\newblock \emph{arXiv preprint arXiv:2305.17144}, 2023.

\end{thebibliography}

\clearpage
\newpage

\appendix

\section{Model Implementation Details}
\label{appendix-implementation}

We implement \vxp with PyTorch~\citep{paszke2019pytorch} and use Python as the programming language of the code~$\code$ and persistent programming environment~$\environ$. To support the wide range of imaging operations required by~$\taskset$, we develop and use a PyTorch library of volumetric medical image utilities, called \textit{Voxel}, available at~\href{https://github.com/dalcalab/voxel}{github.com/dalcalab/voxel}.

\subsection{Language Agent}

We implement the agent model~$\agent$ as a decoder-only transformer stack, using a randomly initialized LLaMA architecture~\citep{touvron2023llama,wolf2019huggingface} with $16$~transformer blocks, a hidden representation of size~$d=512$, a linear representation of size~$\num{2048}$, and $32$~attention heads. We convert text into an embedding space by splitting character groups into tokens (from a vocabulary of size~$\gamma$) and mapping them to a sequence of~$\real^d$ features via an embedding matrix in~$\real^{\gamma,d}$. We use the pre-computed tokenizer released with LLaMA 2, with~$\gamma = \num{32000}$.

The language model auto-regressively generates instruction embeddings~$\instr = \instr^\code \cat \instr^\phi$ based on the input~$\state$. We pass~$\instr^\code$ through a fully-connected layer to obtain text token probabilities~$P(\code)$, and decode into code~$\code$ by choosing the maximum probability token at each sequence position. We pass~$\instr^\phi$ through a fully-connected layer to compute the vision network modulators~$\phi$.

To split~$\instr^\phi$ and~$\instr^\code$, we first transform~$\instr$ embeddings into a sequence of max-probability tokens. We extract~$\instr^\phi$ from all sequence positions that immediately follow special token~\texttt{<MOD>}, and we extract~$\instr^\code$ from all remaining positions. The agent~$\agent$ predicts~\texttt{<MOD>} and subsequent~$\instr^\phi$ features after each volume encoding and generation function argument. We project~$\instr^\phi$ embeddings to~$\phi$ using a fully-connected layer with 32~output channels and SiLU activation.

\subsection{Persistent Environment}

In environment~$\environ$, we predefine a variable corresponding to each volume~$v$. As the code~$\code_i$ is executed, new variables are defined and retained in~$\environ$, persisting across steps. To guide the next instruction step,~$\code_i$ can include \textit{read} operations, which extract the value of a variable in~$\environ$ and embed it in a representation~$z_i$ as feedback in the next state~$\state_{i+1} = \state_i \cat \instr_i \cat z_i$.

For each volume~$v$ passed through~$\enc$, we reduce the spatial dimensions of the deepest layer output using a global max operator. We pass these pooled features through a fully-connected layer to compute~$\varepsilon_v^\circ \in \real^d$. When a \textit{read} instruction is executed on a set of volume encodings~$\encvols$ defined in~$\environ$, we concatenate each~$\varepsilon_v^\circ \in \encvols$ into the feedback embeddings~$z$.

\subsection{Volume Encoder and Generator Subnetworks}

We implement $\enc$ and~$\gen$ as the respective down-sampling and up-sampling arms of a six-level UNet-like model~\citep{ronneberger2015u}. Each level consists of a 3D convolutional layer followed by a latent feature~$\phi$ mixing layer and stream interaction layer with $b = 32$, as defined in Section~\ref{subsec:modeling_details}. All layers use SiLU activations. The spatial outputs at each level are channel-normalized with a group size of four, then max-pooled~($\enc$) or trilinearly upsampled~($\gen$) by a factor of two. Convolution kernels have size~$3^3$, with 32 output channels at the top resolution level and 96 output channels at all other levels.

For all input volume streams, we populate~$\encvols$ with spatial features output at each level in~$\enc$. We use these latent features as skip-connections to corresponding level inputs in the generator~$\gen$, which predicts volumes through a convolutional layer with one output channel. Lastly, we apply the sigmoid activation to generate binary segmentation maps. If multiple volumes from a single scan session are passed as input to \vxp, we compute a merged, session-specific segmentation by extracting the max values across outputs corresponding to each session.

\subsection{Optimization}

We train \vxp using the Adam optimizer~\citep{kingma2014adam} with an initial learning rate of~$10^{-4}$, a batch size of one, and 10 gradient accumulation steps on an NVIDIA A100 GPU. We halve the learning rate after~$10^5$ steps with no improvement in validation accuracy, stopping training after four sets of learning rate updates. We set the volume loss weight~$\lambda = 0.1$.

\section{Training Data Details}
\label{appendix-training-data}

\subsection{Image Preprocessing and Augmentation}

For each image volume, we normalize intensities within the range~$[0,1]$, conform the data layout to a right-anterior-superior~(RAS) orientation, and crop the field of view to a 20 mm margin around the cranial cavity. We co-register all images acquired from each subject using~\citet{hoffmann2024}.

In training, we randomly sample up to 8 (or max available) images corresponding to a scan session. We augment images by applying random affine transformations, spatial intensity distortions (bias field simulations, spatial smoothing,~$k$-space corruptions), exponential scaling, lateral anatomical flipping, cropping, anatomical masking, and voxel resizing. We take advantage of our resolution-agnostic vision network and randomly sparsify training data to reduce voxel throughput and significantly reduce total train time. This random sparsification is performed by sampling slice separations from the range [1, 6] mm or by cropping the field of view. We ensure that the target ROI, if applicable, is not removed during this process. Volume sparsification is performed with 50\% probability for each sample or when total input voxels exceeds a preset threshold to prevent device memory errors.

\subsection{Anatomical Dataset Details}
\label{appendix-anatomy-datasets}

In addition to the pathology datasets outlined in~\ref{images}, we generate segmentations for whole-brain anatomical structures on images from the FSM~\citep{greve2021deep}, OASIS~\citep{marcus2007open, lamontagne2019oasis}, Mind Brain Body~\citep{babayan2019mind}, IBC~\citep{pinho2018individual}, CERMEP~\citep{merida2021cermep}, and Forrest Gump~\citep{hanke2014high} cohorts. We select high-quality acquisitions and thoroughly inspect and correct errors in the label maps. Additionally, we make use of multiple image atlases with precomputed segmentations~\citep{adil2021high,pauli2018high}.

\subsection{Anatomical Structures}
\label{appendix-structures}

We use segmentations of various anatomical classes, listed below. Bilateral brain structures are defined by two distinct hemisphere-specific labels.

Global tissue classes include the brain, dura, skull cavity, cerebrum, cerebral white matter, cerebral cortex, brainstem, cerebellum, ventricular system, and cerebral spinal fluid (CSF).

Brain sub-structure labels include the amygdala, nucleus accumbens, hippocampus, thalamus, caudate, putamen, dorsal striatum, globus pallidus (externus and internus), basal ganglia, hypothalamus, fornix (body, crus, and column), mammillary body, septal nucleus, subthalamic nucleus, habenula, ventral pallidum, extended amygdala, red nucleus, anterior and posterior commissures, pars compacta, pars reticulata, parabrachial pigmented nucleus, ventral tegmental area, fimbria, septum pellucidum, tectum, pineal gland, superior and inferior colliculus, cerebral peduncle, medullary pyramid, medial lemniscus, cerebellar peduncle (superior, middle, inferior), cerebellar gray matter, and cerebellar white matter.

Ventricular sub-structure labels include the lateral ventricle, inferior lateral ventricle, posterior lateral ventricle, anterior lateral ventricle, atrium, third ventricle, fourth ventricle, interventricular foramen, and cerebral aqueduct.

Cortical sub-region labels include the frontal lobe, parietal lobe, temporal lobe, occipital lobe, cingulate cortex, insular cortex, anterior cingulate cortex, caudal anterior cingulate cortex, rostral anterior cingulate cortex, posterior cingulate cortex, isthmus cingulate cortex, frontal pole, middle frontal gyrus, caudal middle frontal gyrus, rostral middle frontal gyrus, superior frontal gyrus, inferior frontal gyrus, pars opercularis, pars orbitalis, pars triangularis, lateral orbitofrontal cortex, medial orbitofrontal cortex, precentral gyrus, paracentral lobule, inferior parietal lobule, superior parietal lobule, supramarginal gyrus, precuneus, postcentral gyrus, entorhinal cortex, fusiform gyrus, parahippocampal gyrus, temporal pole, inferior temporal gyrus, middle temporal gyrus, superior temporal gyrus, transverse temporal gyrus, cuneus, lingual gyrus, and pericalcarine cortex.

\subsection{Radiopaedia Data}
\label{appendix-radiopaedia}

We download and annotate 101 patient cases from Radiopaedia, a radiology reference at \textit{https://radiopaedia.org}. Each case includes text-based notes and scans in the form of 2D image slices. We reconstruct volumetric data by stacking these slices and estimating an affine matrix to map voxel coordinates in world space. We compute this mapping by registering the image to an average template.

\begin{figure*}[!t]
    \centering
    \includegraphics[width=\textwidth]{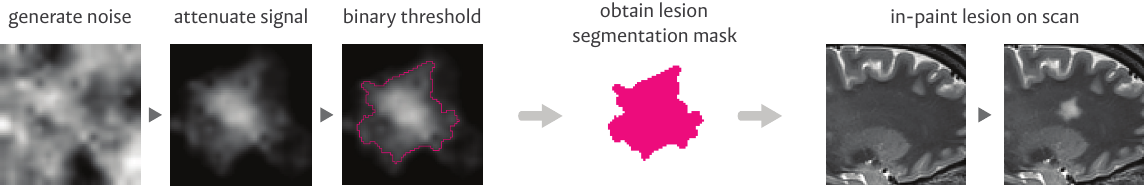}
    \caption{Schematic of the lesion synthesis procedure. A lesion shape is first generated by attenuating and thresholding Brownian noise. The resulting segmentation map is resampled into the target image space, with size and position determined based on anatomical priors. The lesion is in-painted by pasting the tissue mask into the image with procedurally-generated texture and mean signal intensity based on randomly selected relative tissue characteristics.}
    \label{fig:lesion-synth}
\end{figure*}

\subsection{Lesion Synthesis}
\label{appendix-lesion-synth}

To extend the range of pathological features observed during training, we synthesize brain lesions with variable characteristics using a model-based domain randomization technique~\citep{gopinath2024synthetic}. Images spanning diverse acquisition types from $26$ subjects in the OASIS, Mind Brain Body, and CERMEP datasets are paired with whole-brain anatomical segmentation maps as the basis for this process.

For each synthetic lesion, we first sample parameters describing anatomical location, dimensions, intensity relative to surrounding tissue, and texture. As illustrated in Figure~\ref{fig:lesion-synth}, volumetric multi-scale Brownian noise is generated with a signal fall-off matched to the sampled lesion dimensions, then thresholded to produce lobulated structures. These shapes define lesion boundaries, which are further constrained by anatomical maps. For example, parenchymal lesions are restricted to white and gray matter, while ventricular lesions are restricted to cerebrospinal fluid spaces. Lesion interiors are inpainted into the native image using randomly generated textures derived from Perlin noise. The mean intensity of the inpainted lesion is determined by the underlying healthy image signal and sampled relative intensity shift.

We also synthesize multiple lesions per subject with varying properties, providing negative examples for \vxp to learn to differentiate abnormalities based on descriptive features. In addition, we simulate heterogeneous lesions by superimposing secondary Brownian noise–derived masks within an existing lesion, producing intra-lesional components or heterogeneous signal profiles.

\section{Experimental Details and Data}
\label{appendix-experiments}

\subsection{Segmentation Evaluation Data}
\label{appendix-segmentation-data}

In the table below, we summarize the number of unique images from held-out subjects used when comparing \vxp segmentation accuracy to SynthSeg~\citep{billot2023synthseg} and individual specialist models. These evaluations use the same set of anatomical test images (Appendix~\ref{appendix-anatomy-datasets}), so we group them together below.

\begin{table}[H]
\centering
\caption{Numbers of held-out test images and subjects corresponding to the whole brain anatomical segmentation experiment in Section~\ref{subsec:brain-image-analysis} and the multi-task training ablation in Section~\ref{subsec:ablations}.}
\begin{tabular}{lll}
\toprule
Segmentation Target & Images & Subjects \\
\midrule
infarct               & 206     & 206  \\
glioma                & \num{1376}  & 344  \\
edema                 & \num{1386}  & 347  \\
cyst                  & 24      & 11  \\
papilloma             & 16      & 6  \\
meningioma            & 21      & 8  \\
white matter hyperintensities  & 40      & 20  \\
anatomical structures  & 108     & 40  \\
\bottomrule
\end{tabular}
\end{table}

\begin{figure*}[!t]
    \centering
    \includegraphics[width=\textwidth]{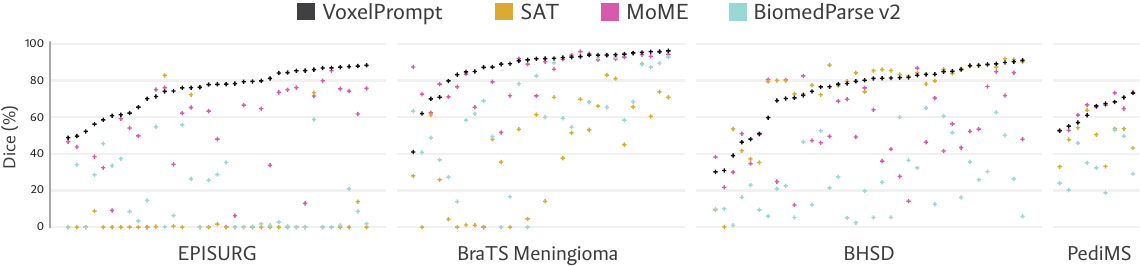}
    \caption{Per-subject accuracy for zero-shot lesion segmentation. Subject indices are sorted along the \textit{x}-axis in ascending order of \vxp prediction accuracy. We find that \vxp consistently achieves high-quality segmentation across all evaluation datasets and lesion types considered.}
    \label{fig:appendix-ranked}
\end{figure*}

\subsection{Zero-Shot Lesion Segmentation Baselines}
\label{appendix-lesion-seg}

\subpara{Baseline settings} For the pathology segmentation experiments, we evaluate baseline performance across diverse input configurations to ensure fairness and avoid bias. For both BiomedParse v2~\citep{zhao2025biomedparsev} and SAT~\citep{zhao2025large}, we explore a range of prompting strategies on each dataset, following the formats recommended in the original repositories or used during their training. We test multiple levels of pathology classification terminology and adopt the phrasing that yields the best performance. BiomedParse v2 does not specify a preferred anatomical orientation, so we evaluate across all possible orientations and report the best-performing one. We also find that BiomedParse v2 does not benefit from resampling inputs to isotropic resolution. The MoME~\citep{zhang2024foundation} baseline requires skull-stripping and affine alignment to the MNI template prior to prediction. We follow  this preprocessing and report MoME's results in the original coordinate system. 

The dataset-specific prompts used as input to language-conditioned methods are described below.

\begin{table}[H]
\scriptsize
\centering
\begin{tabular}{llll}
\toprule
 & VoxelPrompt & BiomedParse V2 & SAT \\
\midrule
BraTS menin. & segment the hyperintense mass & MRI imaging of a brain tumor & meningioma \\
EPISURG & segment the hypointense lesion & MRI imaging of a brain lesion & stroke \\
BHSD & segment the hyperdense lesions & CT imaging of brain lesions & intracranial hemorrhage \\
PediMS & segment hyperintensities & MRI imaging of lesions & white matter hyperintensities \\
\bottomrule
\end{tabular}
\end{table}

\subpara{Additional results} In Figure~\ref{fig:appendix-ranked}, we show per-subject performance plots for all methods on zero-shot lesion segmentation. \vxp consistently outperforms the baseline foundation models for brain pathology segmentation.

\subsection{Scan Geometry Sampling}
\label{appendix-scan-resolutions}

To evaluate the efficiency improvements provided by the native-resolution vision network, we sample test images with spatial geometries drawn from distributions representative of clinical MR and CT brain acquisitions. Image dimensions are sampled uniformly around a $155 \times 190 \times 165$ mm field of view, with a max deviation of $\pm 15$ mm in each dimension. This range is derived from the~$10^\text{th}$ and~$90^\text{th}$ percentile values of image sizes in our preprocessed datasets.

To reflect real-world voxel resolutions, we consider imaging modalities most frequently collected in a standard clinical brain imaging session. For each acquisition type, we define uniform sampling ranges for in-plane resolution and slice separation, based on standard protocol guidelines and empirical resolution distributions observed in our dataset. Voxel spacings are clamped to a minimum of 0.8 mm. During each experimental sample, we randomly select a field of view, acquisition type, and resolution from these distributions, and randomly populate the volumes with Gaussian noise. This distribution, outlined below, is not exhaustive, but is designed to provide a representative coverage of common acquisitions sufficient to evaluate the benefits of native-resolution processing.

\begin{table}[h!]
\centering
\begin{tabular}{lcc}
\toprule
& in-plane spacing (mm) & slice separation (mm) \\
\midrule
T1-weighted (isotropic)   & 0.8 -- 1.2  & iso. \\
T2-weighted               & 0.8 -- 1.0  & 3.0 -- 5.0 \\
FLAIR                     & 0.8 -- 1.0  & 3.0 -- 5.0 \\
diffusion-weighted (DWI)  & 1.5 -- 2.5  & 2.0 -- 3.5 \\
gradient-echo (GRE)       & 0.8 -- 1.2  & 4.0 -- 6.0 \\
perfusion MRI             & 1.5 -- 2.5  & 4.0 -- 6.0 \\
susceptibility-weighted (SWI) & 0.8 -- 1.0  & 1.5 -- 3.0 \\
CT (isotropic)            & 0.8 -- 1.0  & iso. \\
CT (thick slice)          & 0.8 -- 1.0  & 3.0 -- 6.0 \\
\bottomrule
\end{tabular}
\end{table}

\subsection{Image Synthesis for Evaluating Stream Interaction}
\label{appendix-image-synthesis}

Brain image synthesis techniques are used to train neuroimaging models that are robust to acquisition variability and anatomical differences~\citep{gopinath2024synthetic}. These approaches employ domain randomization methods~\citep{tobin2017domain}, in which whole-brain anatomical segmentations are mapped to randomized tissue intensities, warped by spatial transformations, and augmented with simulated artifacts. The resulting synthetic images extend beyond the realistic range of clinical scans, enabling models to generalize across diverse real-world data and tasks~\citep{dey2024learning}.

We adopt this strategy as a controlled framework for evaluating \vxp's ability to integrate complementary information across arbitrary numbers of input volumes. By generating multiple synthetic images from a single anatomical label map, we test whether segmentation accuracy improves as additional inputs are provided.

For stream-interaction evaluation, we employ a standard image synthesis protocol widely used in brain imaging~\citep{gopinath2024synthetic}. Briefly, for each evaluation sample, we sample a whole-brain anatomical segmentation map from a set of OASIS subjects. To generate an individual image from this segmentation, each label is assigned an intensity distribution defined by Gaussian parameters sampled uniformly, following a classical Bayesian segmentation formulation. Voxel labels are then recoded into grayscale values drawn from their respective label distributions, producing synthetic images. Finally, we apply random artifact simulations, including spatial blurring, additive noise, and bias-field distortion.

To generate corrupted images, we synthesize random label maps from multi-scale Brownian noise. Between 10 and 20 noise maps are generated, and voxels are assigned to the index of the maximum-valued map. This synthetic label map is then converted to an image using the same label-to-intensity procedure described above.

\subsection{Stream Interaction Variants}
\label{appendix-interaction-results}

We compare our attention-based stream interaction module with two reduction-based variants commonly used in multi-input medical image analysis \citep{butoi2023universeg}. In these variants, features are aggregated across input streams by mean or max pooling along the stream dimension. The resulting global feature representation is then concatenated channel-wise with the original stream-specific features. This combined feature map is passed through a linear projection layer whose output dimensionality matches that of the block input, ensuring compatibility with the downstream network. We test both mean and max-reduction as alternative aggregation operations to our proposed attention mechanism.

To compare interaction mechanisms, we implement vision models matching the \vxp architecture but without language-conditioning blocks, varying only the stream interaction module. Models are trained on synthetic data (Appendix~\ref{appendix-image-synthesis}) using multi-class Soft Dice loss to segment 35 anatomical brain structures.
At each training step, between one and three images corresponding to a single segmentation target are sampled, with a 10\% probability of including a corrupted image (described in Appendix~\ref{appendix-image-synthesis}).

For evaluation, we generate 500 synthetic image sets, each containing three images derived from a held-out test set of 50 subjects and a predefined subset of corrupted images. We evaluate model performance by varying the number of images provided as input from each set and measuring Dice overlap between predicted and reference segmentations.

\begin{table}[htb]
\centering
\caption{On \textit{uncorrupted} input image sets, attention- and reduction-based stream interaction methods result in similar model segmentation accuracy (Dice), which improves for all methods as image inputs are added for a single forward pass.}
\begin{tabular}{lccc}
\toprule
Method           & 1 image        & 2 images       & 3 images \\
\midrule
attention        & $83.7 \pm 2.8$ & $85.8 \pm 1.8$ & $86.9 \pm 1.4$ \\
max-reduction    & $83.5 \pm 2.6$ & $85.8 \pm 1.8$ & $87.1 \pm 1.5$ \\
mean-reduction   & $83.0 \pm 2.7$ & $85.7 \pm 1.8$ & $86.7 \pm 1.6$ \\
\bottomrule
\end{tabular}
\label{tab:interaction-variants}
\end{table}

\section{Pathology Characterization Experiment}
\label{appendix-classification-results}

\begin{figure*}[!b]
    \centering
    \includegraphics[width=\textwidth]{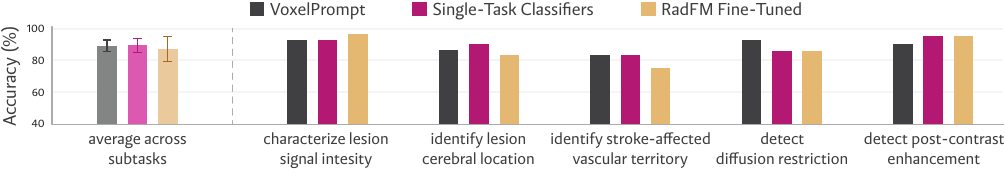}
    \caption{Accuracy of pathology characterization using natural language for five separate classification subtasks. Average subtask accuracy is shown on the left. \vxp (black) parallels the performance of individually-trained, single-task classifiers (purple) and a fine-tuned RadFM model (yellow) -- a state-of-the-art method for 3D visual question-answering.}
    \label{fig:characterization}
\end{figure*}

We evaluate the ability of \vxp to produce a natural language characterization of image features. We focus on five pathology-based visual question-answering tasks (also used during training). These involve classifying lesions based on (1) signal intensity, (2) broad cerebral location, (3) stroke-affected vascular territory, (4) diffusion restriction, and (5) post-contrast enhancement.

\subpara{Data} For each of these tasks, we curate a subset of held-out subjects with relevant features, while ensuring equal representation of possible classification categories in each subset. In total, the evaluation set consists of 102 cases, with per-task breakdowns detailed below. 

\begin{table}[H]
\centering
\begin{tabular}{lll}
\toprule
Classification Task & Images & Subjects \\
\midrule
characterize lesion signal intensity         &  26   & 26 \\
identify lesion cerebral location            &  112  & 30 \\
identify infarct vascular territory           &  16   & 12 \\
detect diffusion restriction                 &  28   & 14 \\
detect post-contrast enhancement             &  40   & 20 \\
\bottomrule
\end{tabular}
\end{table}

\subpara{Evaluation} During evaluation, we consider a prediction as correct if the output natural language response exactly matches the expected characterization. Using a paired \textit{t}-test, we compare the \vxp per-task classification accuracy to that of multiple baselines.

\subpara{Baselines} We compare \vxp to a set of classifier benchmarks, each trained for one of the five pathology characterization tasks in~$\taskset$. As opposed to using language, the single-task benchmark models directly predict label probabilities for a fixed set of task-specific characterizations. We implement these models using the architecture of~$\enc$, with~$\phi$ mixing layers replaced. We reduce the spatial dimensions of the deepest encoder layer output using a global max operator, then compute the maximum value over all input volume streams. To compute classification probabilities for~$n$ possible descriptions, we pass the stream-pooled features to a fully-connected layer with~$n$ output channels and softmax activation. During benchmark optimization, we use the categorical cross-entropy loss on these predicted probabilities.

We also compare \vxp to RadFM~\citep{wu2025towards}, a publicly released, state-of-the-art architecture for medical visual question answering that can process multiple 3D images simultaneously. In our preliminary experiments, we find that the pretrained RadFM cannot generalize to the neuroimaging tasks used in this experiment. Therefore, we \textit{fine-tune} RadFM on our subset of pathology characterization tasks, using the training code released with their pretrained model weights. To fit the optimization within 80 GB of GPU memory, we keep only the first eight hidden transformer layers of the language model and do not modify any other model components. As required by the vision transformer, we resize all input volume spatial dimensions to the nearest multiple of~$32 \times 32 \times 4$. During fine-tuning, we use only the expected language response (without code) as the target text.

\subpara{Results} Figure~\ref{fig:characterization} shows that \vxp achieves an average classification accuracy of~$89.0 \pm 3.6\%$ over all tasks, matching the performance of the single-task benchmarks~($89.3 \pm 4.2$\%) as well as the fine-tuned RadFM model~($87.1 \pm 7.9\%$). These results demonstrate that \vxp can achieve language-based image characterization with comparable performance to specialized classification and medical vision-language architectures, while also able to perform the wide variety of tasks described in the main experiments.

\end{document}